\documentclass[acmlarge]{acmart}
\AtBeginDocument{%
  \providecommand\BibTeX{{%
    \normalfont B\kern-0.5em{\scshape i\kern-0.25em b}\kern-0.8em\TeX}}}

\setcopyright{acmlicensed}
\acmJournal{IMWUT}
\acmYear{2023} \acmVolume{7} \acmNumber{3} \acmArticle{98} \acmMonth{9} \acmPrice{15.00}\acmDOI{10.1145/3610921}

\usepackage{multirow}
\usepackage{graphicx}
\usepackage{subfigure}

\begin{document}


\title[MicroCam: Leveraging Smartphone Microscope Camera for Context-Aware Contact Surface Sensing]{MicroCam: Leveraging Smartphone Microscope Camera for Context-Aware Contact Surface Sensing}

\author{Yongquan Hu}
\orcid{0000-0003-1315-8969}
 \email{yongquan.hu@unsw.edu.au}
\affiliation{%
 \institution{University of New South Wales}
 \country{Australia}}

\author{Hui-Shyong Yeo}
\orcid{0000-0002-9512-828X}
 \email{yeo.hui.shyong@huawei.com}
\affiliation{%
 \institution{Huawei}
 \country{China}}

\author{Mingyue Yuan}
\orcid{0009-0004-5797-0945}
 \email{mingyue.yuan@unsw.edu.au}
\affiliation{%
 \institution{University of New South Wales}
 \country{Australia}}

\author{Haoran Fan}
\orcid{0009-0009-0551-1583}
 \email{haoran.fan@student.unsw.edu.au}
\affiliation{%
 \institution{University of New South Wales}
 \country{Australia}}

\author{Don Samitha Elvitigala}
\orcid{0000-0002-8013-5989}
 \email{s.elvitigala@unsw.edu.au}
\affiliation{%
 \institution{University of New South Wales}
 \country{Australia}}

 \author{Wen Hu}
\orcid{0000-0002-4076-1811}
 \email{wen.hu@unsw.edu.au}
\affiliation{%
 \institution{University of New South Wales}
 \country{Australia}}

 \author{Aaron Quigley}
\orcid{0000-0002-5274-6889}
\email{aquigley@acm.org}
\affiliation{%
 \institution{University of New South Wales}
 \country{Australia}}

\renewcommand{\shortauthors}{Hu et al.}


\begin{abstract} 










The primary focus of this research is the discreet and subtle everyday contact interactions between mobile phones and their surrounding surfaces. Such interactions are anticipated to facilitate mobile context awareness, encompassing aspects such as dispensing medication updates, intelligently switching modes (e.g., silent mode), or initiating commands (e.g., deactivating an alarm). We introduce MicroCam, a contact-based sensing system that employs smartphone IMU data to detect the routine state of phone placement and utilizes a built-in microscope camera to capture intricate surface details. In particular, a natural dataset is collected to acquire authentic surface textures in situ for training and testing. Moreover, we optimize the deep neural network component of the algorithm, based on continual learning, to accurately discriminate between object categories (e.g., tables) and material constituents (e.g., wood). Experimental results highlight the superior accuracy, robustness and generalization of the proposed method. Lastly, we conducted a comprehensive discussion centered on our prototype, encompassing topics such as system performance and potential applications and scenarios.

\end{abstract}




\begin{teaserfigure}
  \centering
  \includegraphics[width=0.9\textwidth]{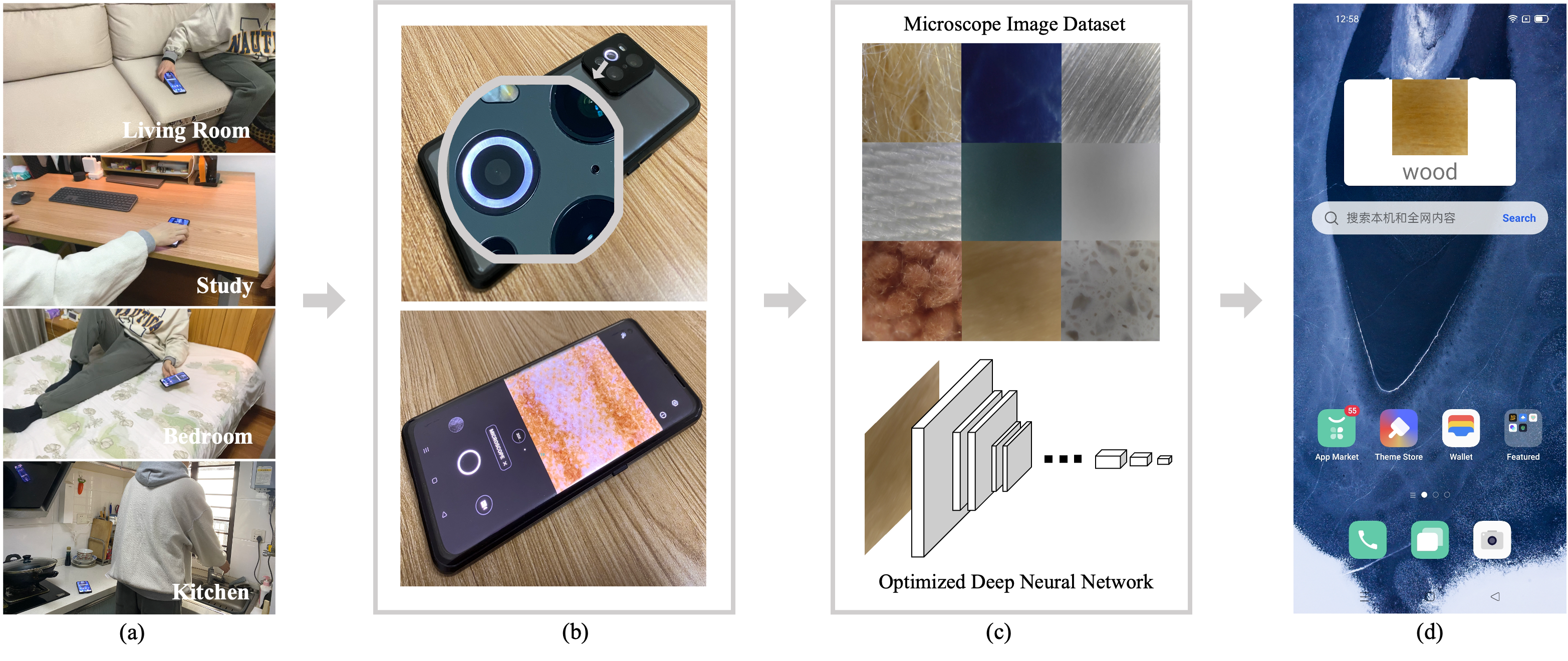}
  \caption{The MicroCam system workflow and use case entail: (a) Placing mobile phones on various surfaces in different scenarios; (b) The upper subfigure displays the phone's rear with an active microscope camera and aperture, while the lower subfigure exhibits the captured microscopic image; (c) Collected images form a dataset (upper subfigure) for training an optimized deep neural network, designed for efficient microscopic surface texture classification (lower subfigure); (d) Surface detection triggers a backend service, launching relevant applications. For instance, if wood is detected, contextual inferences may suggest a study environment.}
  \label{fig:teaser}
\end{teaserfigure}

\maketitle

\section{Introduction}

Advances in pervasive computing, especially mobile-computing, have opened up the potential of context-aware interactions to become part of the fabric of our daily life~\cite{gellersen2002multi,Yeospecam,shi2021fine}. 
However, many nascent forms of context-aware interaction still require extra learning, configuration and user maintenance to achieve the appearance of a seamless and carefree interaction, that many desire. 
An understanding of what the user wishes, as a by product of actions they are performing anyway, remains a challenge. In MicroCam, we design, develop and evaluate a form of mobile interaction, based on the placement action the user is going to perform anyway, through the use of a phone with a microscope camera. 
This work is situated within the wider advancements of positioning and sensing technologies which have afforded new interaction techniques based on context perception. More specifically, within these advancements, a range of discreet and subtle techniques have emerged~\cite{Euanmicrogestures, marquardt2012cross, PohlChartingSubtle}.

Surface and material sensing technology~\cite{HarrisonPlacementAware, yang2012magic,SpectroPhone,Yeospecam,SatoSpecTrans, hwang2013vibrotactor,fujinami2012design,Sensurfaces} offers the potential for more fine-grained mobile phone context awareness compared to other localization techniques reliant on communication signal detection (e.g., GPS~\cite{huang2008improve}, Bluetooth~\cite{1309060}, GSM~\cite{quigley2005proximation}, WiFi~\cite{lee2019random, guo2018accurate, 10.1145/3494954}, or multi-signal fusion~\cite{gellersen2002multi}). In the ``Placement Awareness'' paradigm~\cite{HarrisonPlacementAware}, a crucial facet of context awareness, users gain insight into the specific location of their mobile devices, such as on a bed, a desk, or in their pocket, rather than merely the general vicinity of a bedroom or living room, which may provide further inference regarding the user's current potential behavior. 
Among various sensing categories (e.g., optical~\cite{HarrisonPlacementAware, SpectroPhone,Yeospecam, SatoSpecTrans}, acoustic~\cite{hwang2013vibrotactor}, magnetic~\cite{fujinami2012design}), optical-based sensing methodologies~\cite{HarrisonPlacementAware, SpectroPhone,Yeospecam,SatoSpecTrans} have demonstrated superior performance in differentiating materials and achieving improved recognition outcomes. Despite minor variations in the devices utilized (e.g., SpeCam~\cite{Yeospecam} employs the reflection of a mobile phone's front screen and front camera, while SpectroPhone~\cite{SpectroPhone} depends on the reflection of external LEDs and the rear camera), the fundamental sensing principle is consistent: unique materials yield diverse spectral reflectance. However, spectral discrimination may occasionally lead to information loss. 
In contrast, color images with an original optical base provide more comprehensive features and finer-grained information (e.g., vivid and distinct material textures) compared to reflected spectrum. 
This wealth of information can be harnessed for a wider array of application scenarios, such as recognizing exceptionally small QR code patterns on surfaces, as depicted in Figure \ref{fig:app}, which is unattainable using surface perception based on spectral recognition alone. Consequently, acquiring this enriched information is essential for inferring the semantic context of user behavior in context-aware applications.
Furthermore, with the growing prevalence of macro photography in commercial mobile phones and the advent of devices featuring ``ultra-close macro lenses'' (microscope lenses) (e.g., Oppo Find X3 Pro, Realme GT2 Pro), the potential to capture RGB images of surfaces during camera placement arises. 
In an effort to capture the intricate texture details present in microscopic images, DNNs (Deep Neural Networks) have emerged as highly effective and prevalent methods in recent years \cite{xing2017deep}. However, they also encounter several technical challenges, such as limited training data, substantial computational complexity, and restricted robustness and generalizability. To address these issues, we conducted through a three-pronged approach: (1) collecting a custom dataset for training the network; (2) selecting a comparatively lightweight MobileNet architecture to balance performance and computational complexity; (3) incorporating continual learning into our system to enhance the algorithm's performance.

Meanwhile, beyond the algorithmic performance of the technology itself, we also place considerable emphasis on the practicality~\cite{stephanidis2019seven}, usability~\cite{thuring2007usability,de2006interaction}, and user-friendliness~\cite{trenner1987win,dillon1987psychological} from a user-centered system design perspective \cite{abras2004user,gasson2003human}, aspects that have been limited addressed in previous work. In terms of practicality, accurate scene perception and precise action detection often necessitate the attachment of multiple sensors or devices, such as radar sensors~\cite{YeoRadarCat}, optical sensors~\cite{HarrisonPlacementAware,SpectroPhone}, or multi-spectral sensors~\cite{SatoSpecTrans}. For example, SpectroPhone achieved the inspiring 99\% accuracy for 30 distinct materials using warm and cool white LEDs in conjunction with a smartphone's rear camera~\cite{SpectroPhone}. However, such strategies assume the presence of external hardware support and are incompatible with off-the-shelf devices.
From the perspective of usability and user-friendliness, SpeCam~\cite{Yeospecam}, for instance, can detect surfaces using only the built-in front camera but requires users to place the phone face down with an appropriate bumper case (3mm thickness), which is not a typical orientation for screen usage and obstructs visibility of front display notifications. 
Conversely, methods that capitalize on pre-existing user interactions align more closely with user habits, enabling novel forms of mobile context-aware computing. While the future application potential of this technology is vast, users currently face challenges in readily acquiring the prototype and integrating it into their everyday routines. Overall, our objective is to leverage flexible, consumer-friendly approaches that are easily integrated, user-friendly, and compatible with popular devices, rather than relying on fixed and additional sensors or infrastructures. In this context, we seek to explore and expand the boundaries of interaction based on built-in sensors in existing commercial mobile phones without requiring additional attachments. Secondly, a diverse range of cameras integrated into assorted mobile devices (e.g., GoPro, smartwatches) are also consistently situated on surfaces in various environments, further substantiating the need for camera-based contact sensing. Additionally, the datasets obtained from real-world environments impose minimal constraints on users, which could be expected to further augment the versatility of our interactive technology to enhance the user experience.
Most importantly, sensing detection should occur unobtrusively and be executed automatically without necessitating additional user learning, such as accompanying unintentional behavior, e.g., implicit interaction or subtle interaction~\cite{schmidt2000implicit,ju2015design,serim2019explicating,mueller2020next,PohlChartingSubtle}. Thus, we employ the mobile phone's IMU (Inertial Measurement Unit) data for loop detection, activating the sensing algorithm to capture surface images upon placement states. This quiet, non-disruptive method requires no additional learning costs.


In summary, this paper presents an efficient, lightweight context-aware sensing system utilizing the built-in microscope camera of a mobile device (as shown in Figure \ref{fig:teaser}). Firstly, the IMU sensor is used to monitor the movement status of the mobile phone since people put them anywhere in daily life (Figure \ref{fig:teaser} (a)). Then, once the state of the mobile phone being placed still is detected, the ``microscope camera'' mode is activated to capture microscopic images of surfaces on which the phone is naturally or unintentionally placed (Figure \ref{fig:teaser} (b)). Next, we employ a deep neural network (MobileNet) optimized based on continual learning (Experience Replay method) for object and material recognition (Figure \ref{fig:teaser} (c)). Lastly, some context-aware information can be inferred based on detection results (Figure \ref{fig:teaser} (d)). Specifically, our contributions include:

\begin{itemize}
\item A contact sensing system that integrates built-in hardware (an IMU sensor and a microscope camera of smartphone) and software (an optimized deep neural network). This system detects the placement status of a smartphone and recognizes the surface texture beneath, aiding in mobile context-awareness;
\item An open dataset of surface texture images, featuring object types and material properties, with rich color and detail, obtained through an object-oriented natural acquisition method during everyday smartphone placement activities to improve the generalization of the proposed approach;
\item A improved MobileNet based on continual learning for microscopic images recognition, offering bolstered robustness and reinforced generalization of the algorithm.
\end{itemize}

\section{Related Work} \label{related work}
Our related work spans multiple research areas including context-aware computing, surface and material sensing, in particular for input and interaction, and image recognition based on deep neural networks.

\subsection{Context-Aware Computing} \label{related work_1}
UbiComp (Ubiquitous Computing) represents a post-desktop paradigm of human-computer interaction, in which ``context is any information that can be used to describe the situation of an entity. An entity is a person, place, or object that is considered relevant to the interaction between a user and an application, including the user and application themselves''~\cite{dey2001understanding}. Mobile context-awareness encompasses information such as spatial detail, identity, user details, temporal details, environmental factors, social context, resources, computing factors, physiological data, activities, schedules, and agendas. It aims to provide services that are adaptive, proactive, and automatic, reducing the cognitive burden on users. However, existing work often necessitates explicit interactions, rather than capitalizing on interactions people would naturally undertake.

Pervasive interface usability metrics usually emphasize learnability, efficiency, memorability, error resistance, satisfaction, conciseness, transparency, and invisibility \cite{raychoudhury2013middleware,quigley2010gui}. The final metric, invisibility, assesses the degree to which the interface remains unobtrusive when it could have inferred or deduced the answer. MicroCam exemplifies this principle, harnessing insights gleaned from natural interactions to deduce contextual state information (e.g., activity, location, social status). By merely requiring users to place their phones on various surfaces as they typically would, without necessitating additional learning, MicroCam effectively facilitates mobile context awareness.

The concept of ``placement-awareness''~\cite{HarrisonPlacementAware} emerged as a noteworthy manifestation of context-awareness, demonstrating the value of material identification for both location and activity recognition through placement detection. This approach sought to differentiate materials when a custom-built multispectral optical sensor is placed on various surfaces. Although Harrison et al. acknowledged the significance of material and surface perception based on placement for contextual situation inference, the interactive intent and experience of the ``placement action'' itself are not comprehensively addressed. Moreover, limited by early technological advancements, the prototype is not extensively integrated with mobile devices, leading to the development of a self-contained, lightweight, and portable standalone unit. While this design is versatile, it may not be ideally suited for specific scenarios.

In contrast, the widespread use of mobile phones and wearables today enables the proposed MicroCam system to offer unique application scenarios while permitting implicit interaction with computing systems based on routine activities.

\subsection{Contact Surface and Material Sensing}\label{related work_2}
Contact surface sensing has emerged as a burgeoning research topic, spanning the fields such as UbiComp, HCI (Human-Computer Interaction), IoT (Internet of Things) and robotics \cite{yang2012magic,Phoneprioception,YeoRadarCat,hwang2013vibrotactor,Harmonics,TouchCam}. The capacity to sense contact surfaces or their materials can facilitate a variety of applications, encompassing but not limited to manufacturing, robotic gripping, context-aware computing, and household surface interactions \cite{HouseholdSurfaceInteractions}. 
Microwave, acoustic and optical sensing are the most commonly used methods for surface sensing ~\cite{HarrisonPlacementAware,yang2012magic,SpectroPhone,Yeospecam,SatoSpecTrans, hwang2013vibrotactor,fujinami2012design, YeoRadarCat, lee2019random}.
Microwave sensing of surface materials offers advantages such as the ability to penetrate through various weather conditions and detect objects at long distances with high resolution. Additionally, it can provide valuable data on material properties and subsurface features \cite{YeoRadarCat}. However, disadvantages include signal interference from other sources and potential difficulties in distinguishing between similar materials. Moreover, the technology can be limited by high power requirements and the need for specialized equipment\cite{lee2019random}.
Moreover, acoustic sensing uses sound waves to measure surface properties such as thickness and stiffness~\cite{UltrasonicNondestructive}, and is often used to detect defects in materials. However, it may not be suitable for all surfaces. Furthermore, spectroscopy is highly accurate in identifying a material's chemical composition but can be expensive~\cite{spectroscopysurface}. Meanwhile, optical inspection is a widely used non-destructive method for surface sensing that measures the reflectivity and color of a surface using light-based sensors~\cite{HarrisonPlacementAware,SpectroPhone,Yeospecam,SatoSpecTrans}. It is relatively inexpensive and accessible to a wider range of users. This method can provide highly detailed and accurate information about surface properties, including texture, roughness, and reflectivity.

Our study primarily focuses on optical or visual surface sensing methods. A notable example is the work by Harrison et al. \cite{HarrisonPlacementAware} which utilized a custom-built sensor comprising a photoresistor and light-to-frequency converter, achieving a material accuracy of 94.4\% for 27 placements. SpecTrans \cite{SatoSpecTrans} extended this concept by incorporating multi-spectral sensing, obtaining 99.0\% accuracy even for transparent materials. In addition, Erickson et al. \cite{ClassificationofHouseholdMaterials} employed a spectrometer, achieving a material classification accuracy of 94.6\%. Schrapel et al.'s SpectroPhone \cite{SpectroPhone} identified 30 distinct materials using warm and cool white LEDs in conjunction with a smartphone's rear camera. Conversely, SpeCam \cite{Yeospecam} repurposes a smartphone's built-in sensor, the front camera, alongside the screen as a multi-spectral emitter, facilitating surface sensing with 99.0\% accuracy for 30 materials, while SpectroPhone \cite{SpectroPhone} leverages the rear camera and flashlight to attain comparable results.
However, regardless of whether it's SpecTrans \cite{SatoSpecTrans}, SpectroPhone \cite{SpectroPhone}, or SpecCam \cite{Yeospecam}, the multispectral-based identification approach is primarily suited for classification tasks and struggles to identify patterns on material surfaces. In contrast, RGB-based microscopic image textures offer a more versatile tool, capable of distinguishing between different materials while also capturing finer patterns. Although MagicFinger \cite{yang2012magic}, utilizing microscopic images, has demonstrated impressive test results on 22 textures, its grayscale image input omits certain color information, thus limiting its efficacy for more complex texture classification tasks, such as materials with similar textures but different colors.

Non-vision approaches are also feasible, such as radar \cite{YeoRadarCat}, vibration absorption \cite{hwang2013vibrotactor}, sound echo \cite{Harmonics}, or a fusion of sensors \cite{SurfaceSense, Phoneprioception}. In the literature, Magic Finger by Yang et al. \cite{yang2012magic} bears the closest resemblance to our work. However, their system necessitates a micro camera connected to a sizable optical processing unit, and the camera can only capture black and white images, which limits the available information about surfaces.

\subsection{Deep Learning and Continual Learning}\label{related work_3}
Deep Neural Networks (DNNs) have been extensively employed in a multitude of multimodal tasks, encompassing data analysis, natural language processing, and image processing \cite{deldari2022cocoa,tenney2019bert,targ2016resnet}. The preeminence of Convolutional Neural Networks (CNNs) in computer vision is attributed to their revolutionary advancements compared to traditional machine learning algorithms. Beyond image classification and recognition, researchers have adapted neural networks for additional computer vision tasks such as semantic segmentation, object detection, and video analysis. Specifically pertinent to our research, neural networks have been utilized to recognize surface textures of clothing \cite{ClothingPatternsStearns, RecognizingClothin}, distinct areas of the palm \cite{TouchCam}, and household materials \cite{ClassificationofHouseholdMaterials, MultimodalMaterialClassification, SemiSupervisedHapticMaterialRecognition}.
The groundbreaking work of ResNet \cite{resnet} incorporates residual connections between blocks, enabling the training of exceptionally deep networks and the enhanced ability to learn abstract representations. Recently, efforts have concentrated on rendering deep neural networks lightweight for deployment on mobile devices with limited computational resources while maintaining comparable performance. For instance, MobileNet \cite{howard2017mobilenets} employs depthwise separable convolution, achieving similar performance with fewer parameters than conventional CNN models. We have integrated this lightweight network into our system.

Additionally, continual learning, also known as lifelong learning or incremental learning, is an essential aspect of deep learning that focuses on the ability of a model to adapt to new tasks or knowledge while retaining and utilizing prior knowledge effectively \cite{CML-IOT}. In real-world applications, data is often non-stationary, and new information becomes available over time. Continual learning aims to enable deep learning models to learn from such evolving data streams without suffering from catastrophic forgetting, a phenomenon where the model's performance on previously learned tasks degrades while learning new tasks \cite{Bootstrapping}.
Several strategies have been proposed to tackle the catastrophic forgetting problem in deep learning, such as ER (Experience Replay) \cite{rolnick2019experience}, EWC (Elastic Weight Consolidation) \cite{kirkpatrick2017overcoming}, and Synaptic Intelligence \cite{zenke2017continual}. These methods aim to mitigate the interference between old and new knowledge by either reusing stored past experiences, regularizing the model's weights, or selectively updating the model's parameters.

To sum up, our design, delineated in the subsequent section, derives inspiration from existing research on contact surface and material sensing, context-aware computing, and continual learning for deep learning.



\section{System Design}\label{system_design}

In photography, the concept of ``macro photography'' involves the close-up capture of small subjects, such as flowers, rain droplets or insects. The interest in this form of photograph stems from an interest in what one cannot normally see with the naked eye. Mundane objects can appear magical when inspected in detail and the proximity to the surface of objects reveals a world of hidden textures and details which may delight the eye of the viewer. An alternative term, ``micro photography'', broadly refers to the same concept as ``macro photography''. While there is no unified scientific definition to distinguish these terms, generally speaking, the degree of magnification is the main basis for judgment. Some literature \cite{fuchs2020application, team_2020} suggests that the magnification of macro photography is around 20x or lower, while the magnification of micro photography is typically  higher. While this is not an absolute it does suggest a useful threshold to distinguish the two terms here. 

As a result of this interest, today clip-on macro lens for smartphone are common and cost only a few dollars. Indeed, newer models of smartphone are starting to include built-in macro-lens cameras. Furthermore, such macro camera technology has been advancing and it can now capture surfaces that are so close to the camera, that it is quite literally touching the lens, with a zoom factor up to 30x or even 60x. As a result, such types of macro cameras can broadly be considered as ``microscope cameras'' (e.g., Oppo Find X3 Pro, Realme GT2 Pro) based on the threshold noted previously. It's worth noting that such ``microscope cameras'' are still far from the magnification power of optical microscopes, which range from several 100x to the low  1000x. 
While the magnification factor of 30x to 60x of smartphone is relatively low compared to a lab microscope, it is much larger than an ordinary macro lens, and the sharper texture details and ultra-close shooting distances (only 1-2mm) have made mobile phone microscope cameras interesting. Figure \ref{fig:object} and \ref{fig:material} are example images taken by an actual smartphone with a built-in microscope camera (Oppo Find X3 Pro). Here, we have captured images in minute and fine texture detail of a particular surface. Considering that there are so many details about a surface that are captured by this camera, we see an opportunity in leveraging this ``macro lens'' or ``microscope camera'' to achieve new sensing capabilities for mobile devices. In particular, we are interested in surface sensing, to determine the material of the surface, in a mobile context for input and interaction.

In addition, in terms of algorithm implementation, we chose to adopt deep learning for image identification and classification. Microscopic images can reflect extremely detailed textures of objects, some of which are very close, which requires recognition algorithms with extremely strong resolution and accuracy. With the improvement of computing power, we have witnessed the advancement of deep learning in image classification and recognition in recent years. At the same time, to be more easily deployed on mobile devices with less computing power, some lightweight neural networks have emerged and achieved good performance. Hence, we designed and implemented an approach to recognize the surface, based on these microscopic images, using MobileNet. Also, we incorporate the ER method, a continual learning method, into the foundation of this network to enhance recognition robustness and generalization substantially.

Moreover, our research aims to not only evaluate the technical feasibility and efficiency of surface sensing using the microscope camera but also to improve the user experience of the whole system based on context inference. The design of our system is inspired by how users interact with their smartphone throughout the daily life. For example, we consider the natural placement of a smartphone on a surface while it is not in active use, which occurs many times throughout a day (like when we are in the shower or when the mobile is casually set aside). Consequently, such ``placement actions'' can be characterized as unobtrusive, non-intrusive, and requiring minimal learning. Specifically, the IMU, a common built-in sensor in mobile phones, is frequently employed to determine the motion state of the devices. We propose utilizing the two-dimensional acceleration data (comprising linear acceleration and angular acceleration) to discern the placement state of mobile phones. Upon detection of this state, the microscope camera is invoked for a single instance of surface capture and processing.

\begin{figure}
\centering
  \includegraphics[width=0.86\textwidth]{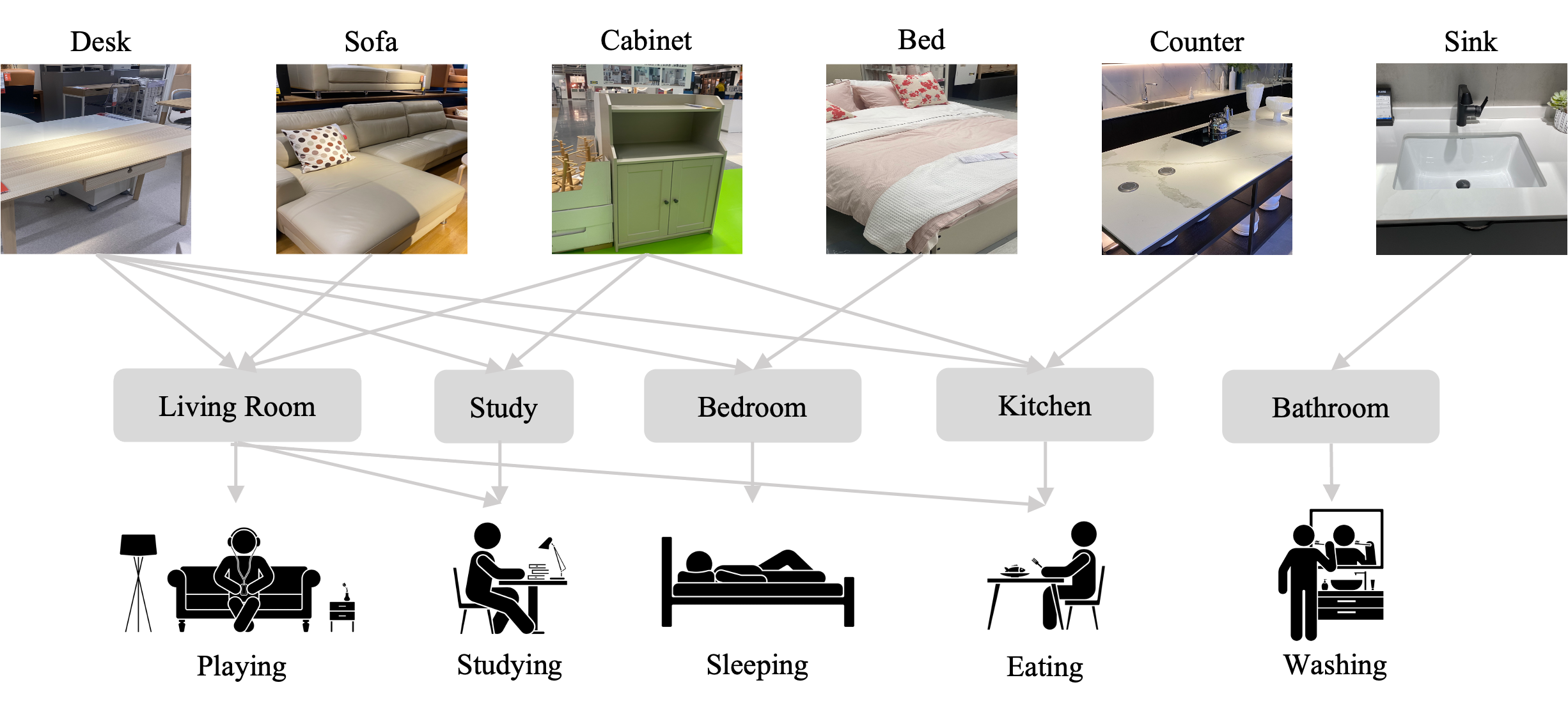}
  \caption{Six objects that people usually place their phones on and some examples of context-awareness including corresponding scenes and activities.}
  ~\label{fig:context_infer}
\end{figure}

Finally, we further consider various real-life scenarios and the locations where the phone will be placed, such as on a desk, on a bed, on a sofa, on the kitchen counter or even the edge of a sink or pool. After several brainstorming sessions and preliminary experiments, we identified six types of objects (bed, desk/table, sofa, cabinet/shelf/closet, sink/pool/bath, counter) where people often place their mobile phones in daily life.
As shown in Figure \ref{fig:context_infer}, these objects often correspond to typical life situations. 
For example, a counter usually signifies a kitchen setting, and a bed typically represents a bedroom, each of which carries implications for contextual awareness.
We discovered that the six common objects found in every household are composed of diverse materials; for instance, the desk in one participant's home might be wooden, whereas in another's, it might be made of fiberboard. Consequently, we required each participant to collect data from all six objects, a task that is easily manageable. However, the types of material varied due to the inherent difficulty and impracticality of requiring participants to gather specific materials. Beyond these criteria, no additional collection requirements are imposed. Participants simply placed their mobile phones on various surfaces as they normally would. This approach, which reflects natural user habits, is referred to as the ``object-oriented'' natural collection method, with the objective of augmenting the generalization of system recognition outcomes.

\section{Dataset}
To develop a neural network adept at discerning distinct surfaces utilizing these image categories, we opted for the independent collection of sample images due to the scarcity of existing datasets in the literature and open-source resources. Notably, outstanding network performance is intimately connected to the quality of the dataset employed.

\subsection{Participants}
We recruited a diverse group of 12 participants (6 males, 6 females), ranging in age from 18 to 39 years (mean = 25.23, SD = 5.2), for the dataset collection. Participants are designated P1 to P12 and represented various backgrounds: six participants are academic students from different disciplines, including computer science (2), mechanical engineering (1), mathematics (1), medical and health (1), and art and design (1); the remaining six participants are office workers from various industries, such as law (1), finance (1), accounting (1), UI/UX (2), and telecommunications (1). All participants had at least two years of experience using smartphones in their daily lives.

\subsection{Task and Procedures}
In order to acquire authentic data from real-life situations, we invited the 12 participants to record their daily mobile phone usage. As previously mentioned in section \ref{system_design}, we identified six common object types on which mobile phones are typically placed in everyday life, which are closely related to various living environments such as the living room, bedroom, kitchen, study, and bathroom. This ``object-oriented'' approach to naturalistic data collection focuses on these six object categories.

To begin with, we presented participants with a comprehensive overview of the experimental procedure and demonstrated the appropriate use of the designated smartphone (Oppo Find X3 Pro). Subsequently, each participant is instructed to utilize the provided phone as they normally would for a period of three days, ensuring that they placed the device on the predetermined objects at least once per day to capture as diverse a range of surfaces as possible. While participants are encouraged to collect surface images at three distinct intervals throughout the day (morning, noon, and evening), this is not a strict requirement. Considering the availability of only one smartphone of the specified model, the device is allocated to each participant (P1 to P12) in a sequential manner, resulting in the whole data collection process spanning approximately one month.

\begin{figure}
\centering
\includegraphics[width=0.92\textwidth]{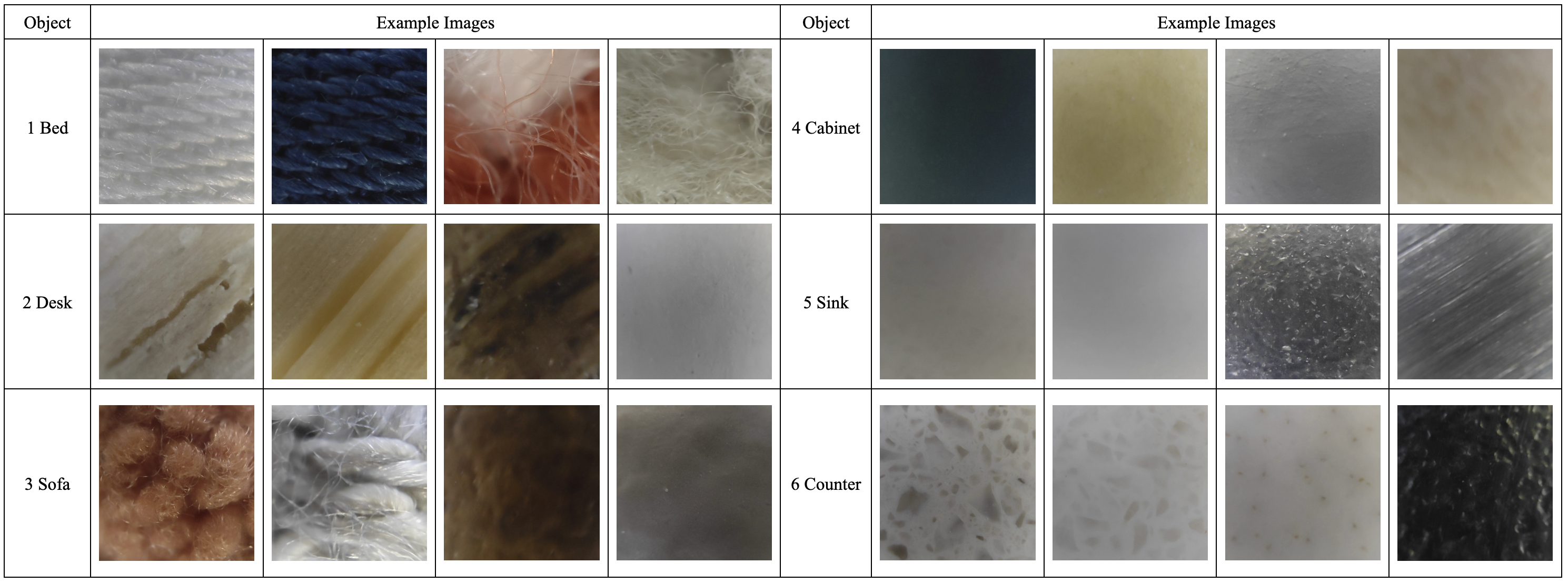}
  \caption{Some example images of 6 types of objects.}
  ~\label{fig:object}
\end{figure}

\begin{figure}
\centering
  \includegraphics[width=0.86\textwidth]{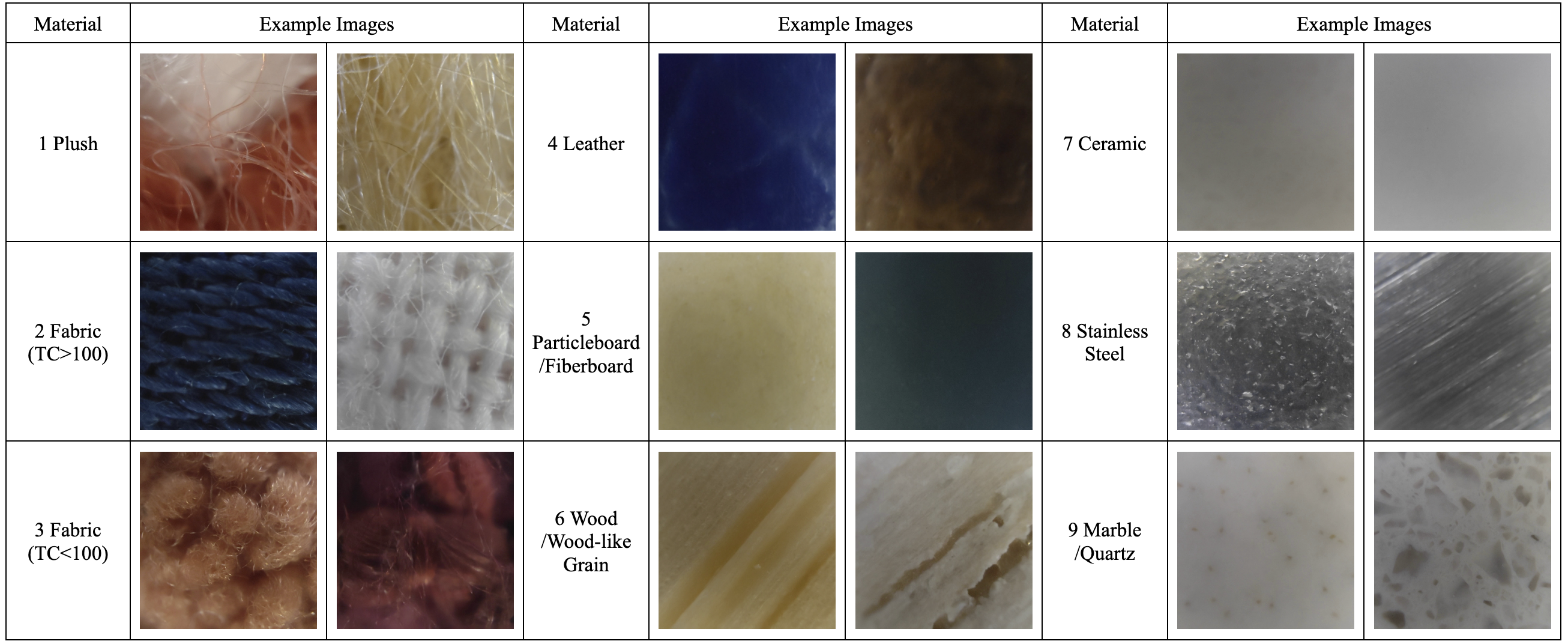}
  \caption{Some example images of 9 types of materials (TC:Thread Count).}
  ~\label{fig:material}
\end{figure}

It is crucial to highlight that although minimum placement requirements are set, there is no imposed upper limit; thus, the aforementioned six objects (bed, desk/table, sofa, cabinet/shelf/closet, sink/pool/bath, counter) are likely to be collected multiple times per day for each participant, with varying frequencies of repetition. During the recording process, participants are advised to randomly move and rotate their phones. Apart from these considerations, the entire use process remains unsupervised and uninterrupted, in line with the user's natural habits of using smartphones (such as how often the phone is placed on the surface, the scene where it is placed, etc.).
Following data collection, the gathered videos are sampled at a rate of three frames per second to avoid extracting excessively similar images, and some images are discarded due to substantial motion blur (e.g., when the recording involved overly rapid rotation) that resulted in excessive distortion. Ultimately, the accumulated surface images are classified into 6 object categories (as shown in Figure \ref{fig:object}) and their corresponding 9 material properties (as shown in Figure \ref{fig:material}). Subsequent sections offer further details.

In short, through our natural, object-oriented data collection approach, the amassed surface images exhibit substantial diversity, encompassing a lengthy time span (over one month), various shooting angles (achieved by rotating the phone), and distinct lighting conditions (corresponding to different times of day). We posit that this diversity contributes to a dataset that more closely aligns with real-world conditions, thereby enhancing the generalization capabilities of the proposed system.

\begin{center}
\begin{table}
  \centering
  \caption{The mapping relationship between object types and material types in our dataset.}~\label{tab:mapping_relationship}
  \begin{tabular}{c|c}
  
    {\small\textit{Object Types}}
    & {\small\textit{Material Types}} \\
    \midrule

    \multirow{2}*{1 Bed} & {1 Plush} \\
    & {2 Fabric (TC>100)} \\
    \hline
    \multirow{2}*{2 Desk/Table} & {5 Fiberboard/Particleboard} \\
    & {6 Wood/Wood-like Grain} \\
    \hline
    \multirow{2}*{3 Sofa} & {3 Fabric (TC<100)} \\
    & {4 Leather} \\
    \hline
    \multirow{2}*{4 Cabinet/Shelf/Closet} & {5 Fiberboard/Particleboard} \\
    & {6 Wood/Wood-like Grain} \\
    \hline
    \multirow{2}*{5 Sink/Pool/Bath} & {7 Ceramic} \\
    & {8 Stainless Steel} \\
    \hline
    {6 Counter} & {8 Marble/Quartz} \\

  \end{tabular}

\end{table}
\end{center}

\subsection{Data Statistics and Description}

Figure~\ref{fig:object} and \ref{fig:material} showcase examples of images captured using the microscope camera of the mobile device. It is important to note that the images for the surfaces of the six objects and nine materials are identical, with the distinction being in their respective property labels. Alternatively, it can be understood that the ``material'' label serves as a supplementary descriptive attribute to the ``object'' label. For instance, ``sofa'' is an object label, and in the data we collected, it could be either ``fabric'' or ``leather''. Consequently, both terms function as material labels, as they further specify the material composition of the sofa.

Specifically, surfaces categorized by object include: 1) Bed; 2) Desk/Table; 3) Sofa; 4) Cabinet/Shelf/Closet; 5) Sink/Pool/Bath; 6) Counter, while surfaces categorized by material consist of: 1) Plush; 2) Fabric (TC>100); 3) Fabric (TC<100); 4) Leather; 5) Fiberboard/Particleboard; 6) Wood/Wood-like Grain; 7) Ceramic; 8) Stainless Steel, 9) Marble/Quartz. The term ``TC'' is an abbreviation for Thread Count, which measures the number of threads woven into one square inch of fabric and is often utilized to describe fabric density. In actual tests, we discovered that both ``sofa'' and ``bed'' contain the material ``fabric''. Assuming we classify their ``fabric'' as the same material, even if our algorithm can accurately identify its material, it will be incredibly challenging to further discern whether it is a ``sofa'' or a ``bed''. We envision our material classification to be as precise and unique as possible, meaning that the same material frequently corresponds to only one object, but in reality, the same object may encompass multiple materials. Based on this observation, we employ TC as a metric to differentiate ``sofa fabric'' and ``bed fabric'' into two categories. For the purpose of this study, we determined 100 to be a suitable threshold (the physical size range of our square microscopic image in Figure~\ref{fig:object} and \ref{fig:material} is approximately 3 $mm^2$).
The density of bed fabrics is typically higher than that of sofa fabrics, which serves as the key factor for the neural network to distinguish between the two categories.
Moreover, the mapping relationship between objects and materials is presented in Table \ref{tab:mapping_relationship} in detail.

We amassed a total of 35,284 images in this study. Furthermore, approximately 3,000 images are collected for each individual participant. For the six object types, each type had a minimum of 4,809 images and a maximum of 7,207, with approximately 12-16 surface cases per object type. Regarding the nine material types, each type had at least 2,014 images and no more than 6,850, with the number of corresponding surface cases for each material type ranging from 7 to 13. Additionally, our dataset includes metadata about the surface (type labels) and the folder is structured according to the ``person-object-material'' hierarchy. For example, ``person 2-sofa-leather'' refers to the leather sofa collected by the second participant. It should be emphasized that since our data collection is based on an object-oriented approach, each participant can gather data on six types of common objects; however, it is not guaranteed that all nine types of material data can be collected. This is consistent with our common understanding, as not all homes or companies will contain all these common objects due to material diversity. Consequently, participants are not required to deliberately search for specific materials but rather use the objects as they typically would. This further underscores our characterization of this data collection method as ``natural'', requiring no additional learning or effort costs.

To summarize, utilizing this data collection methodology enables us to tackle some of the overfitting problems frequently observed in neural networks. The heterogeneity of the collected data, encompassing diverse participants, lighting conditions, time intervals, and angles, along with the sparse sampling density (3 fps), collectively assists in addressing these issues.


\section{System Implementation}
\subsection{Hardware Configuration}\label{hardware_config}
We employed the OPPO Find X3 Pro smartphone (256GB storage, 4500mAh and 8GB RAM) and its two built-in sensors in our study. The first, the IMU sensor, is a common component in smartphones that measures linear and angular motion utilizing accelerometers and gyroscopes. It offers data on device orientation, acceleration, and rotation for a variety of applications. In our research, we analyzed two-dimensional linear and angular acceleration data to ascertain whether the phone is situated on a surface.
The second sensor, the microscope camera (magnification 30x or 60x), is a distinctive feature of this phone model. It permits users to capture close-up images at up to 30x magnification from extremely close distances. The camera also features a ring light (as shown in Figure \ref{fig:teaser} (b) upper subfigure) encircling the lens, ensuring consistent brightness for texture capture. Its short focal length (approximately 1mm) enables users to simply raise the phone's back using a standard case, facilitating effortless placement on surfaces for clear microscopic images (as shown in Figure \ref{fig:teaser} (b) lower subfigure) without suspending the device in mid-air to achieve focus.  In the prototype configuration, we employed a transparent phone case with a thickness of 1mm to maintain the requisite sensing distance.
In addition, all deep neural network training and testing is done on an Alienware X17 R2 laptop. The laptop's configurations are as follows: 1TB SSD, 32GB Memory, 12th Gen Intel Core i9-12900H CPU, NVIDIA GeForce RTX 3070 Ti Laptop GPU, and Windows 11 Home operating system. The pre-training time of MobileNet and ResNet we built on this laptop is about 3 hours and 8 hours respectively.

\begin{figure*}
\centering
  \includegraphics[width=1\textwidth]{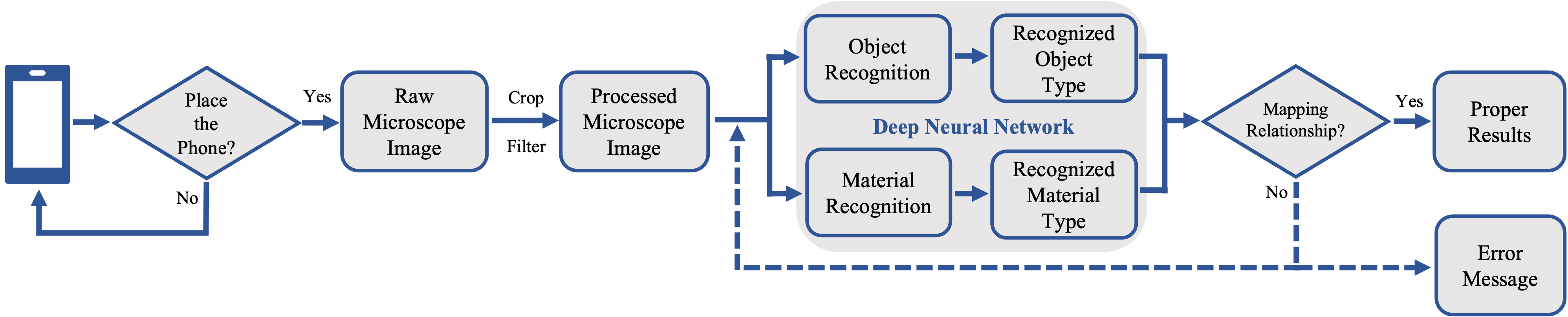}
  \caption{ The schematic representation of the MicroCam system pipeline.}
  ~\label{fig:MicroCam}
\end{figure*}

\subsection{Software Algorithm}
\subsubsection{System Pipeline}

As depicted in the Figure~\ref{fig:MicroCam}, we initially develop a simple Android program to leverage the IMU, allowing the mobile phone to unobtrusively capture surface images in a stationary horizontal state for automatic context awareness. Specifically, we assess the following three aspects: (1) Real-time IMU data monitoring with fixed thresholds for LA (Linear Acceleration) from the accelerometer and AA (Angular Acceleration) from the gyroscope; when both absolute values simultaneously fall below their respective thresholds, the mobile phone is considered stationary horizontal state (denoted as S1); otherwise, it is deemed in another state (denoted as S2); (2) Surface image sampling is activated only when the mobile phone transitions from S2 to S1; (3) Repeated image sampling activation is avoided while the mobile phone remains stationary, and when the static duration exceeds a certain TT (Time Threshold), the program shifts from foreground to background processing, resuming foreground operation upon reactivation. In our practical tests, the triaxial thresholds for LA and AA are respectively set at [0.04, 0.04, 0.04] $(m/s^2)$ and [0.02, 0.02, 0.02] $(^\circ/s)$, while the TT is established at 30 $(s)$. It should be emphasized that the values of these thresholds may vary with different models of mobile phones and different test environments.

Secondly, once surface acquisition is activated, the raw microscopic images collected are processed by cropping, and subsequently filtering out exceedingly blurry images based on the LoG (Laplacian of Gaussian) method \cite{bansal2016blur}. The Laplacian of Gaussian (LoG) method is a technique for edge detection and feature extraction in image processing. It combines Gaussian smoothing to reduce noise and Laplacian edge detection to identify intensity changes, which is widely used for IQA (image quality assessment). By distinguishing true edges from noise artifacts, LoG effectively detects edges in noisy images while preserving important details. In short, this procedure results in the elimination of 31 images. Following this, the processed microscopic images are utilized as inputs for the neural network on a high-end PC during the training and inference phases.

Then, we implement the deep neural network part of the algorithm using the PyTorch framework \cite{NEURIPS2019_9015}. 
Although the highest available resolution is 1920*1080, for the purpose of reducing the amount of calculation and speeding up the processing, the inputs to the network are normalized and resized to 3*224*224. Each input image comprises three channels of RGB, with each channel featuring a 224*224 two-dimensional spatial resolution. During training stage, the training images are augmented with horizontal flip, rotation and random shift. We set the batch size to 16 and use an Adam optimizer with a learning rate of 0.0001 and trained the network for 20 epochs. The trained models are saved and later loaded during real-time user testing. Furthermore, it should be emphasized that we build two identical parallel networks for two different tasks, namely: (1) object recognition (object classification); (2) material recognition (material classification). For object classification, we want to identify the object where the phone is placed on, out of the 6 possible objects, and use softmax function at the last layer for 6 outputs. For material classification, we want to identify the material of the object out of 9 possible classes, and use the same architecture, whereas we employ a softmax function at the last layer for 9 outputs. 

Finally, it is worthwhile mentioning that we add a ``validation'' step after the network outputs the prediction results (as shown in the diamond box on the right in the Figure~\ref{fig:MicroCam}). According to Table \ref{tab:mapping_relationship}, if the predicted results satisfy the mapping relationship between objects and materials, 
they would be output; if not, the recognition fails, and then our program could try the recognition again or terminate the recognition with an error message (in the case of multiple re-recognition failures). Through this way, some obviously wrong predictions can be ruled out.

\begin{figure*}
\centering
  \includegraphics[width=0.8\textwidth]{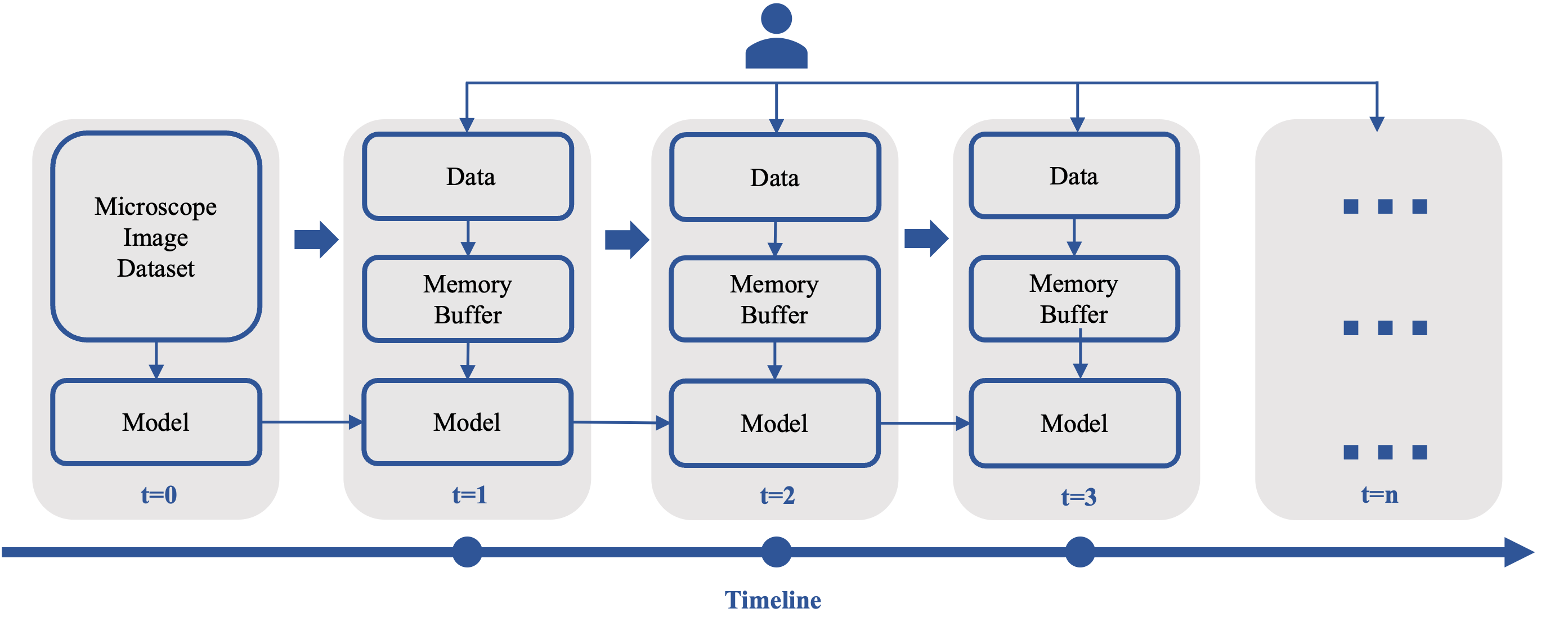}
  \caption{ The continuous learning (ER method) process with time changing for MicroCam system.}
  ~\label{fig:CL}
\end{figure*}

\subsubsection{Continual Learning}\label{cl_er method}

We notice that although deep DNNs could achieve satisfactory accuracy on established datasets, they struggle to maintain good performance when faced with continuous input of new data in practical applications. A typical issue is ``catastrophic forgetting'' for DNNs, where learning a new task (new distribution) leads to forgetting experiences from previous tasks (old distribution), resulting in performance degradation. CL (Continual Learning) is proposed to address this issue, but as it remains an ongoing challenge, we introduced a basic CL method: ER (Experience Replay) \cite{rolnick2019experience}, to preliminarily enhance MicroCam's performance in practical applications.
ER is a method designed to mitigate the issue of catastrophic forgetting by storing and reusing past experiences in memory buffers. The memory buffer, as the core component of the experience replay technique, is responsible for intelligent decision-making and model updating. As depicted in Figure \ref{fig:CL}, we assume that the network model has been trained well on our established microscopic image dataset at time t=0. At subsequent time steps t=1, 2, ..., new data samples are continually inputted. These samples consist of surface images automatically captured by the mobile phone and labels (ground truth) manually annotated by the user. Subsequently, the memory buffer adjusts the new input data and original data according to a predetermined strategy, such as extracting portions of both original and new data into the buffer and assigning them different training weights as per the strategy. The data in the memory buffer is then utilized to update the model parameters to optimize output performance.

Nonetheless, ER is dependent on meticulous design and fine-tuning to achieve exceptional results. In our experiments, we pre-test the performance of memory buffers with sizes of 120, 240 and 500 during the training process and ultimately selected 500 as the optimal buffer size. Furthermore, we also employ five optimization tricks to enhance experience replay in continual learning, drawing upon existing research \cite{buzzega2021rethinking}. The five tricks are as follows: (1) Independent Buffer Augmentation: This trick applies data augmentation techniques individually to each task-specific buffer instead of applying them jointly to the entire buffer, which increases the diversity and robustness of the replay experience, preventing overfitting and forgetting; (2) Bias Control: A strategy that modifies the neural network classification by adding new output units and adjusting old output units when learning new categories. The adjustment is performed by computing a bias correction term based on the statistics of the old and new categories. This approach balances classification scores between old and new classes, preventing classifier bias towards new classes; (3) Exponential Learning Rate Decay: Employing an exponential schedule with a smaller decay factor enables faster learning of new tasks by decaying the learning rate. This trick could effectively prevent the network from overwriting old knowledge with new knowledge by reducing weight update magnitudes; (4) Balanced Reservoir Sampling: This method samples experiences from the memory buffer in a balanced manner, ensuring each category has an equal probability of being selected, thus preventing class imbalances in memory buffers. However, this strategy may cause certain classes to be underrepresented or overrepresented during replay; (5) Loss-Aware Reservoir Sampling: A trick for extracting experiences from memory buffers using loss values as a measure of difficulty or information content, which is effective for replaying experiences with more challenging or correlated samples, improving learning efficiency.

In summary, through these strategies, the five aspects corresponding to continual learning — policy, frequency, batch, enhancement, and regularization — have all been improved in our MicroCam system. Moreover, we contend that ER is a promising method for mitigating catastrophic forgetting and enhancing forward transfer in continual learning scenarios, but there remains substantial room for improvement.


\section{Evaluation}

We conducted a comprehensive evaluation of MicroCam from several perspectives. Firstly, we briefly compare MobileNet and ResNet-50. Although we currently deploy deep neural networks on high-end PCs, we anticipate future transplantation to mobile devices for independent operation, which will enhance the system's portability and wearability, while balancing performance and computational requirements. 
Meanwhile, we compare to the results of MagicFinger \cite{yang2012magic}. 
Secondly, we performed a more detailed assessment of the lightweight MobileNet, including cross-validation and confusion matrix analysis. Subsequently, we present the performance optimization achieved through continual learning. Finally, we provide test results for systematic power consumption and latency to further demonstrate the system's merits.

\subsection{Study 1: Performance Comparison}\label{study1}

Firstly, it is noted that some analogous efforts are undertaken in prior research, such as MagicFinger \cite{yang2012magic}, which employs microscopic images for surface classification and identification purposes based on an ultra-compact macro camera strapped to a finger. A significant distinction in our methodology is the utilization of color microscopic images as system input, in contrast to MagicFinger, which relies on grayscale images. Accordingly, we regard the classification of grayscale images mentioned in MagicFinger as a baseline, present the result comparison for analyzing the performance differences between color and grayscale images as input.

Secondly, we conduct a comparative performance analysis between ResNet-50~\cite{resnet}, a well-established deep CNN architecture, and MobileNet-v2 (version 2), a lightweight architecture. We utilize these two networks as feature extractors for both object and material classification tasks. In this expeditious comparison, the dataset is partitioned by randomly selecting 1/10 of the data as the test set, while the remaining data constitutes the training set. We employ top-1 classification accuracy as the evaluation metric.


    

\begin{table}[H]
  \centering
  \caption{Performance comparison of ResNet-50 and MobileNet-v2 and a comparative analysis employing grayscale (baseline) and RGB images as inputs.}~\label{tab:resNet-50 and mobileNet}
  \begin{tabular}{l r r r r r}
    {\small\textit{Model}}
    & {\small \textit{Weights(M)}}
      & {\small \textit{FLOPs(G)}}
      & {\small \textit{Input Image Color Mode}}
       & {\small \textit{Object Top-1 Acc(\%)}}
        & {\small \textit{Material Top-1 Acc(\%)}}\\
    \hline

    \multirow{2}{*}{ResNet-50} & \multirow{2}{*}{25.56} & \multirow{2}{*}{4.14} & Grayscale       & 97.12 & 98.33\\
                               &                        &                       & RGB & 99.30 & 99.47\\
     \midrule[0.1pt]

    \multirow{2}{*}{MobileNet-v2} & \multirow{2}{*}{1.7}   & \multirow{2}{*}{0.59} & Grayscale       & 96.70 & 97.58\\
                               &                        &                       & RGB & 98.23 & 99.15\\
  \end{tabular}
\end{table}

The performance comparison results are compiled in Table \ref{tab:resNet-50 and mobileNet}. It is evident that classification outcomes utilizing RGB microscopic images as input surpass those using grayscale images, indicating the color RGB microscopic images contribute positively to object and material classification accuracy. This can be attributed to the richer information contained in RGB, which facilitates the network's ability to identify challenging examples. For instance, some sink and cabinet examples in Figure \ref{fig:object} exhibit similar textures but different colors, making it difficult to distinguish using grayscale images. Additionally, the material classification accuracy exceeds that of object classification, as some object categories encompass multiple materials, complicating the recognition process. For instance, the object category ``table/desk'', as illustrated in Table \ref{tab:mapping_relationship}, corresponds to two distinct material textures, namely ``fiberboard/particleboard'' and ``wood/wood-like grain''.

Upon comparing the performance disparities between the two networks under the same color mode input, MobileNet demonstrates advantages in both performance and complexity. For example, with RGB microscopic image input, MobileNet-v2 achieves object and material classification accuracies of $98.23\%$ and $99.15\%$, respectively, representing a decrease of $1.07\%$ and $0.32\%$ compared to ResNet-50. These results highlight that MobileNet attains comparable accuracy to ResNet-50 while maintaining merely 1/7 of its complexity. In conclusion, we select MobileNet-v2 as our primary prototype network and utilize color RGB microscopic images as input. Building upon these foundations, we proceed to deliver a more in-depth evaluation and a detailed optimization in the following sections.

\subsection{Study 2: Cross-Validation}\label{study2}
\subsubsection{Cross-Validation Results}
Following a comparative analysis of network architectures, we opt for MobileNet-v2 and conducted additional evaluations. In this section, we exclusively present the results of MobileNet-v2. 
The outcomes of the cross-validated tests are illustrated in Table \ref{tab:2 cross-validation}. We partition the images into training and test sets using two distinct methods: time-based and person-based splitting.

\begin{table}[H]
  \centering
  \caption{Cross-validation classification performance of MobileNet-v2 (RGB microscopic image inputs).}~\label{tab:2 cross-validation}
  \begin{tabular}{l r r}
    {\small\textit{Classification}}
    & {\small \textit{Time-based Split Acc(\%)}}
      & {\small \textit{Leave-1-Person-Out Split Acc (\%)}}\\
    \midrule

    Object & 98.44 & 95.56\\
    Material & 99.25 & 96.96\\
  \end{tabular}
\end{table}


In the time-based split, we divide data by collection time into 10 parts, and then apply 10-fold cross-validation to evaluate. 
The results of mean accuracy are $98.44\%$ (SD = 0.17) (object) and $99.25\%$ (SD = 0.39) (material). The accuracy of both object and material is very high, and the fluctuation of each test result is also quite small (The value of the two SDs is extremely small). Such a a situation is actually to be expected, because the test set considerably overlaps the training set. The way we solve overfitting for this kind of evaluation is to only take 3 frames per second.

For the person-based split method, we segregate the training and test sets according to each participant. A total of 12 participants are involved, with each individual's data serving as the test set in rotation, while the remaining 11 individuals' data composed the training set. Given that each person constitutes a basic unit, variations in usage habits among individuals result in differing objects and materials upon which the mobile is placed, as well as diverse test durations, lighting conditions, and shooting angles. Consequently, the training and test set data diverge. We argue that this approach mitigates overfitting and data leakage issues. Nonetheless, our results demonstrate a commendable average accuracy of $95.56\%$ (SD = 5.97) (object) and $96.96\%$ (SD = 2.82) (material). Analyzing the outcomes, the leave-1-person-out split method exhibits marginally lower performance than the time-based test, as the training set lacks the test case.

Additionally, we observe that material-based classification tasks are simpler than object-based tasks, yielding $0.81\%$ (time-based split) and $1.40\%$ (leave-1-person-out split) higher performance. This can be attributed to certain objects utilizing similar or identical materials. For instance, tables and cabinets may employ the same wood material, rendering the object classification task more challenging than that of material classification.

\subsubsection{Confusion Matrix Analysis}

We generate confusion matrices under two evaluation methodologies for six objects (1 bed; 2 desk/table; 3 sofa; 4 cabinet/shelf/closet; 5 sink/pool/bath; 6 counter) and nine materials (1 plush; 2 fabric (TC>100); 3 fabric (TC<100); 4 leather; 5 fiberboard/particleboard; 6 wood/wood-like grain; 7 ceramic; 8 stainless steel; 9 marble/quartz).

\begin{figure}
\centering
\subfigure[]{\includegraphics[width=5.6cm]{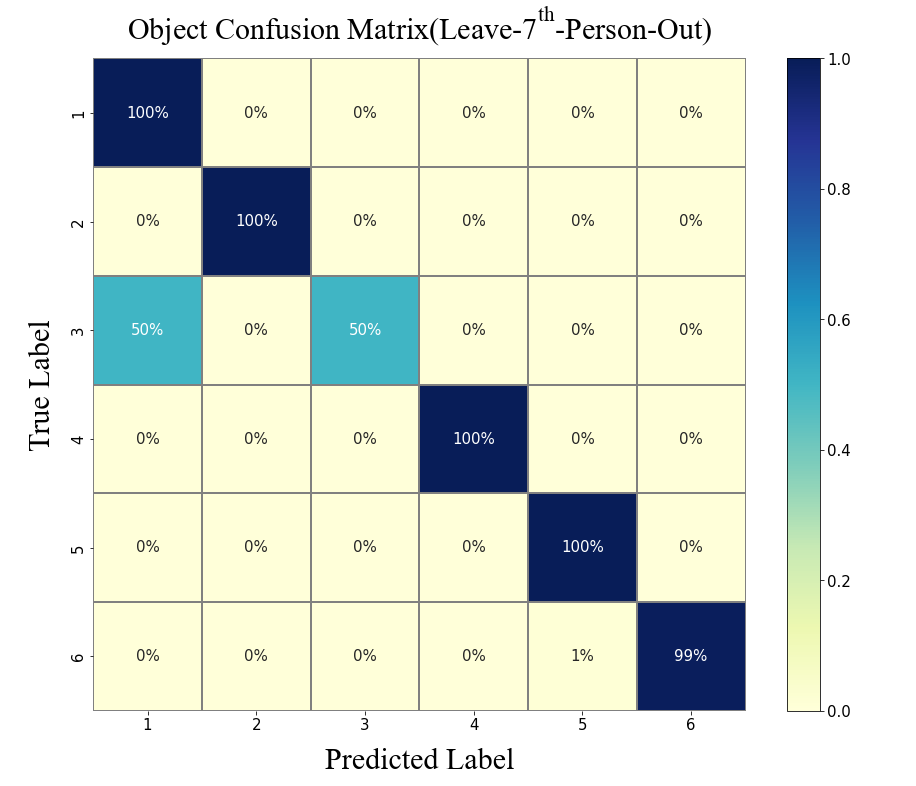}} 
\subfigure[]{\includegraphics[width=5.6cm]{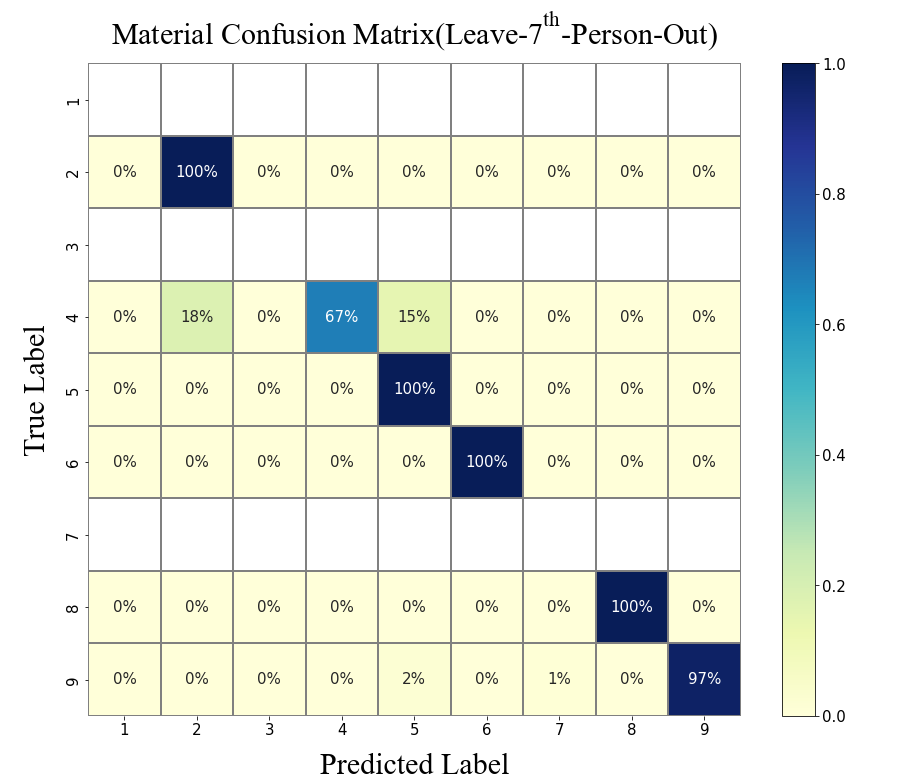}}
\\ 
\caption{Object and Material confusion matrix for leave-7th-person-out when using MobileNet-v2.}~\label{fig:matrix_person7} 
\end{figure}

Prior to examine the overall confusion matrix, it is valuable to focus on a single individual's confusion matrix to clarify the underlying calculations executed, such as the test results pertaining to the $7^{th}$ participant, as a representative example. Figure \ref{fig:matrix_person7} displays the object and material confusion matrices using a ``leave-$7^{th}$-person-out'' evaluation. In other words, the data gathered by the $7^{th}$ person constitutes the test set, while the data obtained by all other participants forms the training set. This ``leave-one-person-out'' approach is applied to each of the 12 participants in our study.

\begin{figure}
\centering
\subfigure[]{\includegraphics[width=5.6cm]{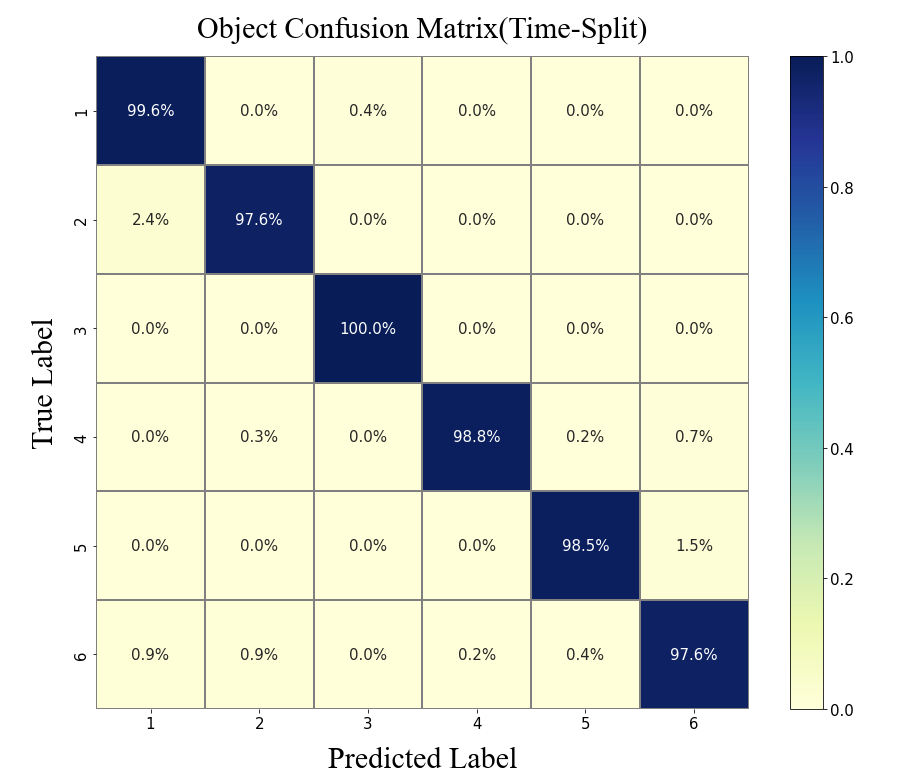}} 
\subfigure[]{\includegraphics[width=5.6cm]{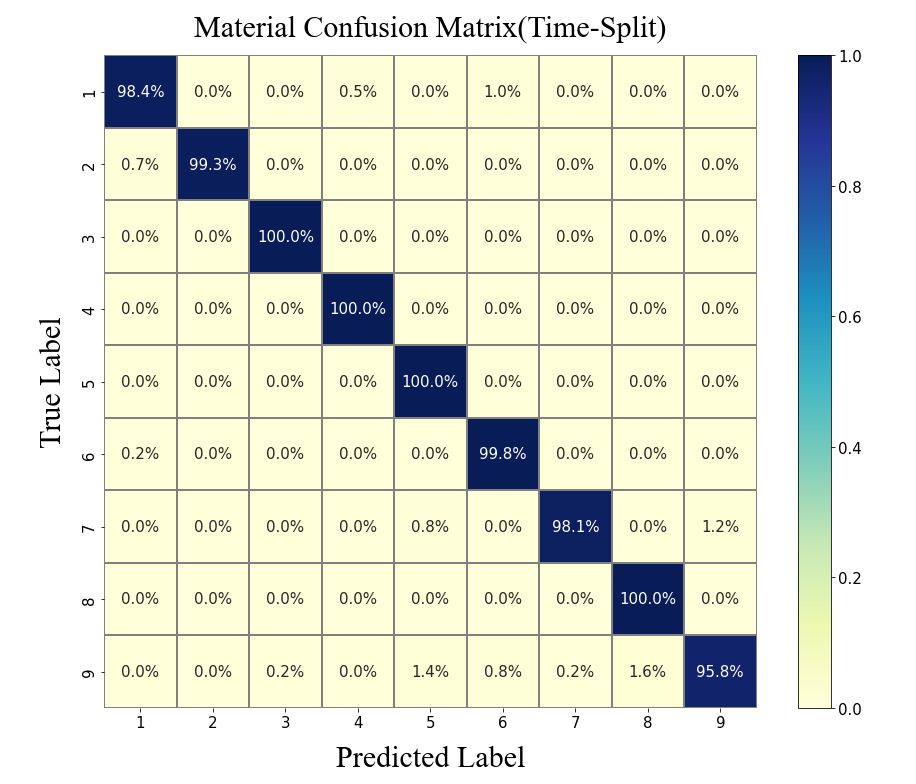}}
\\ 
\centering
\subfigure[]{\includegraphics[width=5.6cm]{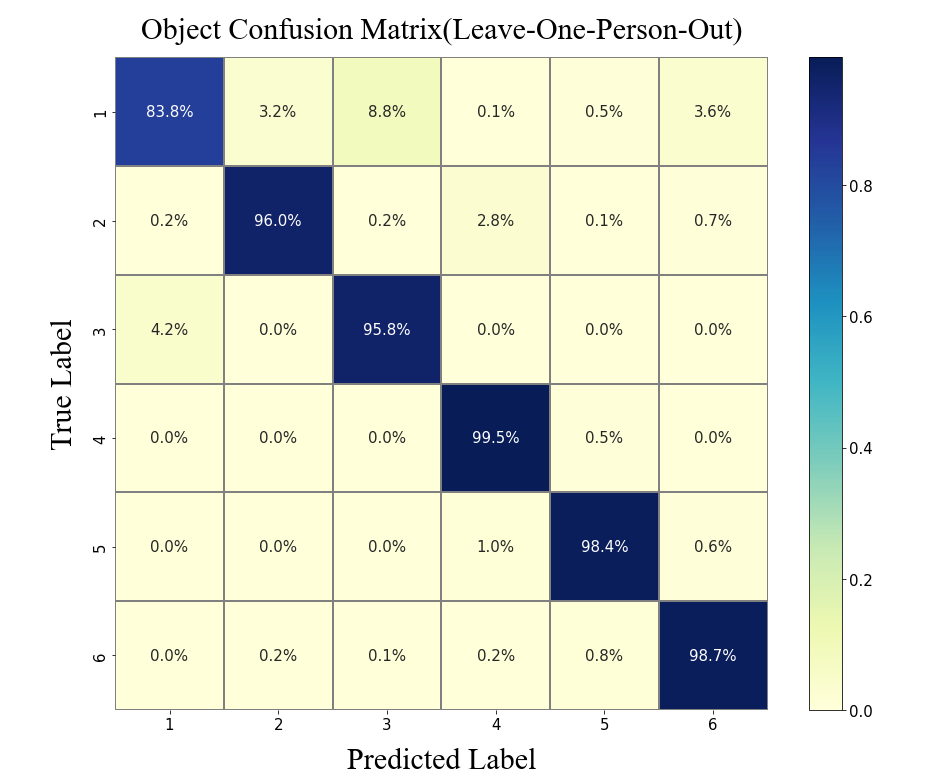}}
\subfigure[]{\includegraphics[width=5.6cm]{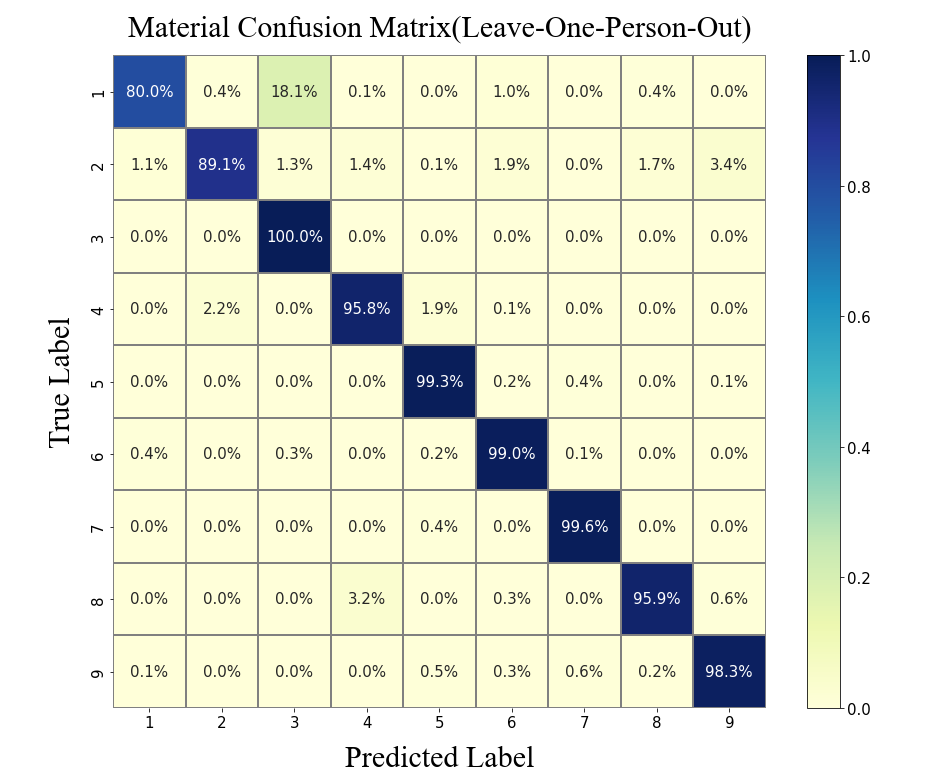}}
\caption{Object and Material confusion matrix for two types of cross-validation of MobileNet-v2. (a) \& (c) Object classification; (b) \& (d) Material classification.}~\label{fig:matrix} 
\end{figure}

Each column represents the ground truth, and each row represents the result of model prediction. As we mentioned previously, when collecting data, we require each participant to collect at least 6 kinds of objects once a day. However, it is not guaranteed to cover all 9 kinds of materials, which is also the origin of our collection method, namely ``object-orientation''. When calculating the confusion matrix, for the leave-one-person-out evaluation method, a material may be missing from the material confusion matrix of a certain participant. For example, in Figure \ref{fig:matrix_person7} (b), the 1, 3 and 7 types of material are not collected by the $7^{th}$ participant. Therefore, when calculating the total confusion matrix for the leave-one-person evaluation, if a person lacks a certain material category, that person would not be counted in the calculation of the average for that category.

For analysis, in Figure \ref{fig:matrix_person7} (a), we can see that the ``1 bed'' and ``3 sofa'' classes are often confused, and 50\% of the sofas are mistakenly identified as beds, which is due to the similarity of the materials. The collected data is limited, and as a result, the model still does not generalize well enough for these two object labels. 
For Figure \ref{fig:matrix_person7} (b), it can be seen that 4 types of leather are misclassified, 18\% are incorrectly identified as category 2 (TC>100), and 15\% are incorrectly identified as category 5 (wood board). We checked these misclassifications and found that these occurred when the color of the image we tested is too dark or too light. This occurs when the image is blurred, or the surface texture is not obvious, which can result in a misclassification. With our current prototype, users should ensure the device is placed in a stable manner to avoid such misclassification problems. This however, is a matter for future work as discussed in section \ref{future_work}. However, other categories can be well distinguished, indicating that there are still obvious differences in the surface texture between different material categories.

Another example can be seen in Table \ref{tab:mapping_relationship}, where both of the objects of class 2 (desk) and class 4 (cabinet) correspond to two materials: class 5 (particleboard/fiberboard) and class 6 (wood/wood-like grain). We originally think that these two classes would be easily misclassified by the neural networks, however this approach continues to perform well. One possible explanation is that for our actual collection, the desks with wood grain predominate, while most of the cabinets/shelves/closets are made of particleboard. As a result, in most cases, they remain straightfoward to distinguish.

For the overall confusion matrix in Figure \ref{fig:matrix}, we note that, under the evaluation of time-based split, both object and material recognition have high accuracy, as one might expect. Instead, let's focus more on the leave-one-person-out evaluation results. As shown in Figure \ref{fig:matrix} (c), in object recognition, 8.8\% of category 1 (bed) is misidentified as category 3 (sofa), and 4.2\% of category 3 (sofa) is misidentified as category 1 (bed). As noted previously, this is due to their similar material composition.
For material classification in Figure \ref{fig:matrix} (d), 18.1\% of class 1 (plush) are mistaken for class 3 (fabric (TC>100)). We can see from Figure \ref{fig:material} that they are indeed similar and therefore easily confused. We suggest that in practical applications, we can further improve the discrimination of the two classes by increasing the amount of training data for this class. 
In addition, other categories can be well distinguished, indicating that the surface textures of different object and material categories contain obvious differences.
Finally, for the categories that are easily misclassified, we need to pay special attention in the application, and improve the accuracy through the retraining or adaptation of the model to the user data.


In summary, due to the promising results of leave-1-person-out split method, future work should investigate a larger training data to demonstrate the generalization performance of the model. 

\subsection{Study 3: Continual Learning}\label{study3}

\begin{figure}
\centering
\includegraphics[width=\textwidth]{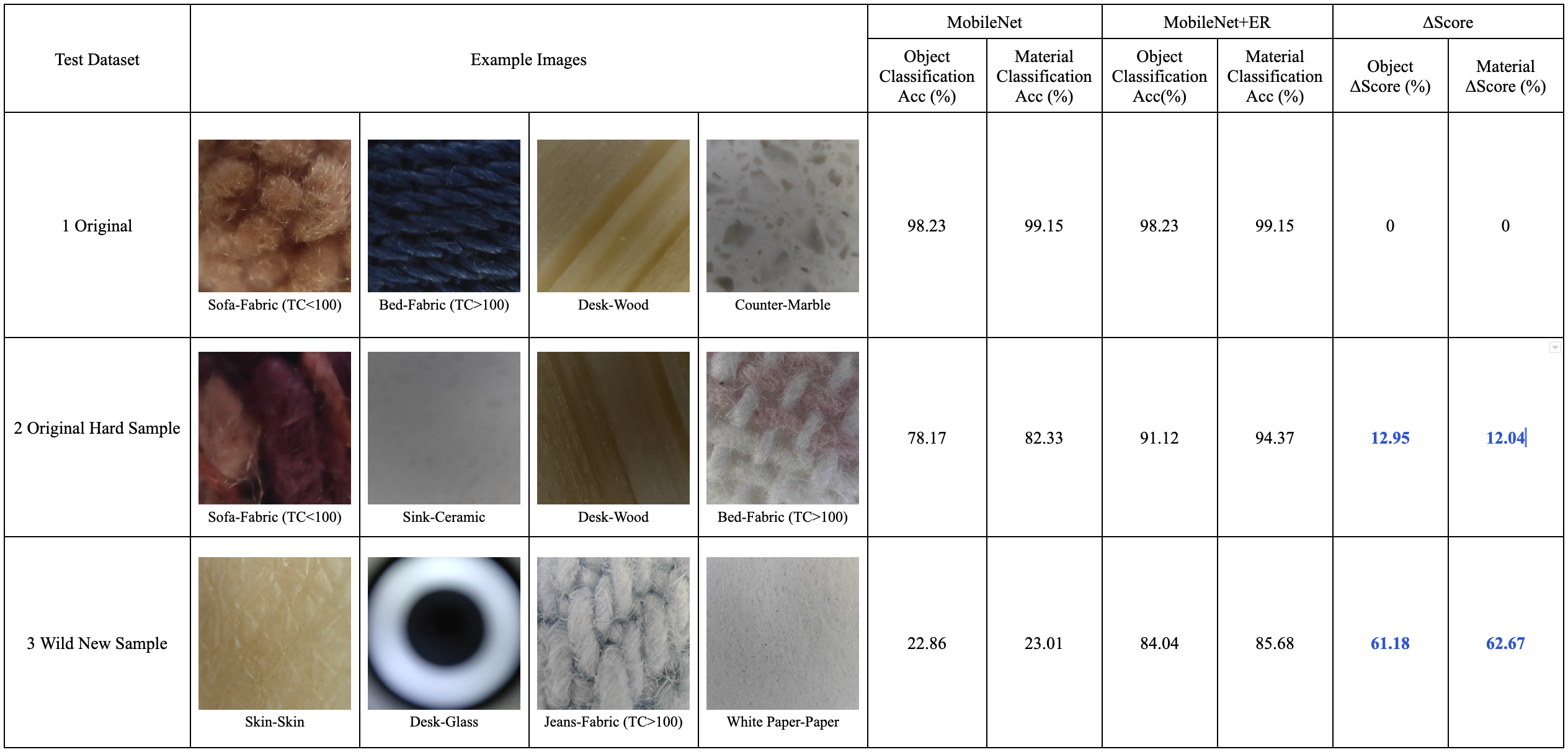}
  \caption{Performance comparison of integrating continual learning (ER method) into MicroCam. Example image captions follow a two-level label naming convention: object type-material type. For instance, ``Desk-Wood'' denotes a wooden desk.}
  ~\label{fig:cl_performance}
\end{figure}

As outlined in section \ref{cl_er method}, we employ the continual learning ER method to optimize the MobileNet-based algorithm of MicroCam. Figure \ref{fig:cl_performance} presents a comparison of MobileNet and MobileNet+ER performance across three test datasets, utilizing the testing methodology described in section \ref{study1}. Specifically, the three test datasets comprise: (1) 3528 images included, obtained by randomly sampling 1/10 of the original dataset; (2) 3000 images included, consisting of difficult instances with low recognition accuracy—some selected from the original dataset and others manually altered with interference processing to increase difficulty. For instance, the four example images in Figure \ref{fig:cl_performance} exhibit reduced recognition rates due to lens defocus, negligible texture, artificially lowered image brightness, and increased noise; (3) 1000 images included, constituting newly acquired samples from real-world scenarios. Concerning the labels of these new sample images, two situations arise: part of sub-labels (object type/material type) of the two-level labels can be found in the original dataset, such as ``Desk-Glass'' and ``Jeans-Fabric (TC>100)''; all sub-labels  are entirely new and not included in the original dataset, such as ``Skin-Skin'' and ``White Paper-Paper''.

From the results depicted in Figure \ref{fig:cl_performance}, the ER method does not yield improvement in the network performance on the original dataset, as this process has yet to involve the memory buffer update. However, for the second test dataset, the performance gains for object and material classification are $12.95\%$ and $12.04\%$, respectively. Given that the data in the second test dataset originates from the original dataset without the introduction of new label categories, this outcome demonstrates that the ER method enhances the robustness of MicroCam. Moreover, the evaluation on the third dataset exhibits a substantial boost of $61.18\%$ and $62.27\%$, respectively. It is attributed to the fact that the supervised learning approach based on the neural network is unable to identify newly added categories; however, upon incorporating ER, the model is capable of continually learning new experiences while retaining the previously acquired knowledge. In this process, we employ various techniques to mitigate catastrophic forgetting and optimize recognition outcomes. The presence of new label categories in this dataset suggests a considerable improvement in the algorithm's generalization. 
In summary, through the incorporation of the continuous learning ER method and the implementation of corresponding optimization techniques, we substantially augment the algorithmic robustness and generalization capabilities of the MicroCam system.

\subsection{Study 4: System Overhead}\label{study4}
To gain a more comprehensive understanding of the load situations of our MicroCam system in practical applications, we conducted tests focusing on two aspects: energy consumption and the response latency.

\subsubsection{Energy Consumption}

Considering that we currently possess only a single mobile phone with an integrated microscope camera, the OPPO Find X3 Pro, we have conducted power consumption tests for the MicroCam system solely on this device. Employing the BatteryHistorian power monitoring tool \cite{BatteryHistorian}, we measured energy consumption and reported the average from an experiment involving 100 repeated inferences on the smartphone. It is important to note that, as previously mentioned, we presume that placing mobile phones on various surfaces represents typical user behavior in everyday life, which also aligns with the initial design intent of MicroCam. Consequently, here term ``every time'' running the program refers to the complete process of setting down the phone, capturing a microscopic image, and subsequently processing it.
Based on our findings, the OPPO Find X3 Pro can support average 120 microscopic image acquisitions per $1\%$ of battery power consumed by the smartphone for the MicroCam system. This outcome suggests that the power consumption of MicroCam is comparable to that of the standard camera mode on the mobile phone.



\subsubsection{Response Latency}

Response latency is a critical element influencing user experience in smartphone applications. The delay primarily consists of two components: the duration required for the Android application to capture a microscopic image and the time needed for MobileNet to process the image. The former entails real-time monitoring of the mobile phone's motion status through loop detection of IMU data, followed by microscopic image sampling upon determining the phone's static state. The average duration for this process is approximately 2.1 seconds (calculated from 50 trials using the developed Android application). The latter refers to the inference time of the MobileNet network and the output of results, which averages less than 0.1 seconds (based on 100 trials using MobileNet on an Alienware X17 R2 laptop). The laptop's specifications could be found in section \ref{hardware_config}.

\section{Findings and Discussion}\label{discussion}

\subsection{Discussion 1: Study Methods and Performance}
In our dataset, as people moved and rotated the phone along the surface while collecting data, and we only extracted 3 frames per second, the images are different. What's more, each of the 12 participants used the phone for three days (spanning day and night) and their usage habits varied. All of these reduce the overfitting and enhance the generalization of our findings.

The average object and  material recognition rates are 98.44\textasciitilde99.25\% when using Time-Split and 95.56\textasciitilde96.96\% when using leave-one-person-out evaluation. 
The latter is more realistic because the testing dataset is never seen during training. This is done to evaluate the robustness and generalization of the system on unknown surfaces. Since such a limited number of surfaces are collected, for example, only 5 people provided relevant data for class 1:plush in the material classification task. In the future, with a larger training dataset (e.g., the size of ImageNet), we believe that the results can be improved and generalized to wider, real-life surfaces.


However, collecting a very large dataset of surfaces is non-trivial. We realized that for general users, there can be infinite number of different surface textures. 

Thus, to address this issue, we incorporate the concept of continuous learning to bolster the robustness and generalizability of the algorithm. Enhanced robustness guarantees improved performance even in the face of challenging sample recognition, while amplified generalizability ensures the model's capacity for learning and adaptation each time new data is introduced, thereby offering users the opportunity for data and label customization.

We experiment with different neural network architectures, a complex and a lightweight architecture, where the results only differ by a little. This suggests that it is sufficient to use a lightweight neural network architecture (e.g., MobileNet, EfficientNet) to strike a balance between computing speed and accuracy, considering that the system should run in a smartphone itself. Furthermore, we incorporated continuous learning (ER method) to bolster the robustness and generalization of the algorithm, thereby significantly augmenting the practicality of the MicroCam system when encountering a wider array of data in real-world settings.

In addition, an IMU-based sensor fusion approach for horizontal stationary state detection of mobile phones further improves the deployment performance of MicroCam. It ensures a seamless integration of data sampling, image processing, and context awareness, functioning as a prime example of implicit interaction. Consequently, it facilitates a non-intrusive, low learning curve, and personalized user experience, optimized for user satisfaction.

\subsection{Discussion 2: Potential Applications and Scenarios} \label{applications}

\begin{figure}
\centering
  \includegraphics[width=1\textwidth]{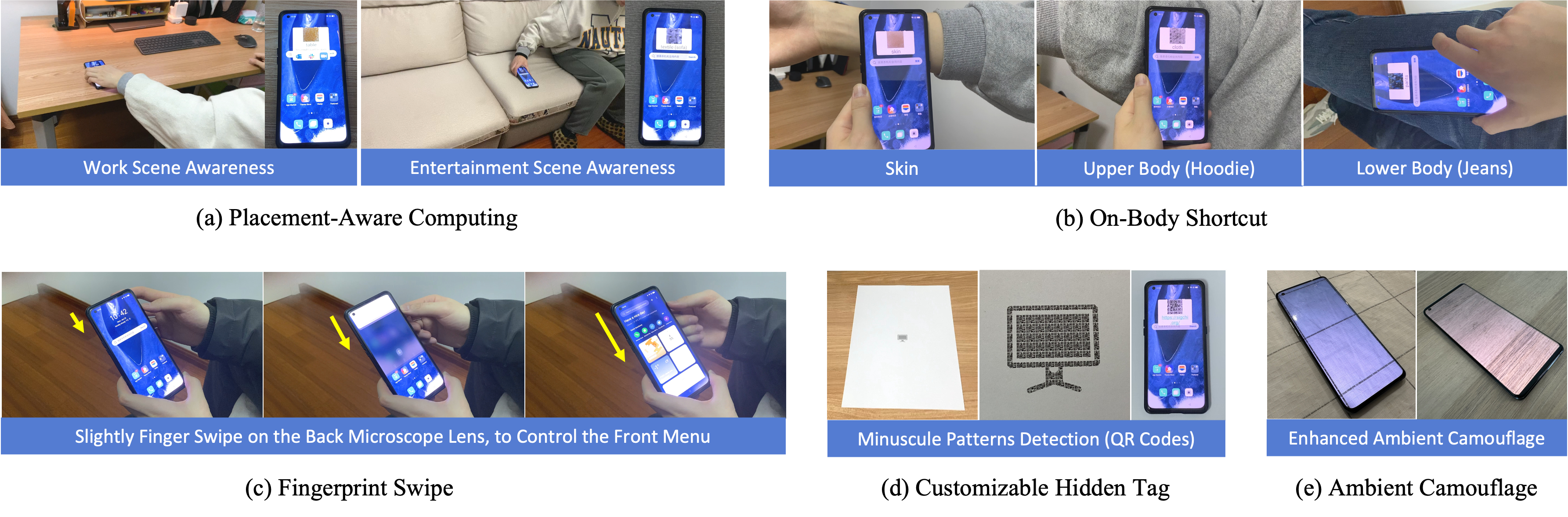}
  \caption{Examples of five potential applications for MicroCam.}
  ~\label{fig:app}
\end{figure}

Leveraging MicroCam surface sensing technology, we present a range of applications aimed at improving the overall user experience, as illustrated in Figure \ref{fig:app}. To provide a detailed overview of these applications, we have devised a template encompassing the following aspects: application description, benefits, supported scenarios, target users, required computational/setup resources, and interpretation of the sketch shown in Figure \ref{fig:app}.

\subsubsection{Placement-Aware Computing}\label{placement_aware}

As mentioned in section \ref{related work_1}, a core and pervasive application of context-aware computational MicroCam systems is leveraging surface sensing results for contextually informed decision-making. Specifically, for MicroCam, we envision its ability to discern the surface it is situated on and autonomously adjust the phone's settings accordingly, such as transitioning to silent mode or providing personalized recommendations. The advantages of such applications are evident: seamless adaptation to the user's environment without necessitating explicit actions, enhancing user experience, and minimizing interruptions in diverse situations. These applications can support an extensive array of scenarios, including home and work settings like offices or bedrooms. To implement this application, in addition to the MicroCam system, further algorithms and hardware based on data mining might be required, which could involve recording private information (user habits, browsing history, etc.) for tailored adaptation. In Figure \ref{fig:app} (a), we present two examples: the first detects a phone placed on a desk for an extended period, inferring that the user may be working, and consequently sets the phone to silent mode; the second example recognizes a living room sofa and automatically displays news or game suggestions for entertainment purposes.

\subsubsection{On-Body Shortcut}\label{onbody_shortcut}

The recognition outcomes of the MicroCam system can serve as a trigger for body shortcuts. By touching the phone to various body parts (arms, abdomen, legs, etc.), it can initiate specific commands or launch particular applications \cite{MnemonicalBodyShortcuts}. This is akin to scanning an NFC tag with your phone, but without the actual hardware tag. Distinct body parts can represent different actions or meanings. For instance, touching the abdomen might indicate that the user is hungry and wishes to order food, prompting the system to launch the food ordering app. Similarly, touching the pants could signify that the user is on the move and desires to initiate a map navigation application. A comparable application is proposed in RadarCat~\cite{YeoRadarCat}. However, RadarCat necessitates supplementary hardware on top of standard mobile phones. In contrast, MicroCam utilizes the built-in macro camera of the smartphone. Moreover, considering the wealth of information provided by microscopy, more refined body shortcuts can be realized, such as estimating skin moisture content from microscopic images and further inferring human body conditions. Consequently, this application is well-suited for individuals seeking swift and convenient support from mobile devices. We suggest that, based on these advantages, the application can be employed more frequently in mobile scenarios and support auxiliary functions. During development, additional body recognition algorithms (e.g., gesture or posture recognition) and corresponding applications can be incorporated to extend the application's functionality. In the example depicted in Figure \ref{fig:app} (b), for preliminary demonstration purposes, we have simply set different response modes for the skin, upper body (hoodie) and lower body (jeans): vibration and two rings with distinct rhythms. The demonstration effect can be observed in the supporting demo video. More diverse and varied response methods can be added in subsequent development stages.

\subsubsection{Fingerprint Swipe}\label{fingerprint_swipe}

Fingerprint swipe is proposed to facilitate microscope-based finger sensing. We envision that, under this application, the information obtained by the microscope can be utilized in two aspects: firstly, the detection of fingerprint texture, where more detailed fingerprint information can be employed to enhance the security of the device's authentication; secondly, detecting minute finger movements as a form of interactive input. Due to the microscope's high magnification, it is highly sensitive to extremely small movements, which can be harnessed for subtle and discreet interactions \cite{PohlChartingSubtle} as well as one-handed use. The aforementioned features are well-suited for privacy requirements in public places, enhancing the system's usability, privacy, and security. For instance, users can unobtrusively interact with their mobile phones in public spaces to initiate customized phone operations. Some complementary technologies for discreet and small interactions are considered for use with this application. In the example provided in Figure \ref{fig:app} (c), we implemented one of the application's functions, which is to determine the direction of finger movement based on the background subtraction method \cite{piccardi2004background}, subsequently triggering the downward slide of the front menu bar.

\subsubsection{Customizable Hidden Tag} \label{customizable_tag}

MicroCam can decipher custom tags embedded in everyday surfaces that are too minuscule to be detected by the naked eye, subsequently triggering specific actions or launching applications. For instance, micro-watermarks can also be concealed within various images and even texts, utilizing our encryption-oriented technology. Another example includes hiding images within human faces, thereby achieving nested images or mosaic photos. This method is characterized by its high concealment and security, enabling the concealment of interaction with physical objects and facilitating subtle or implicit interactions. The potential application scenarios for this method are vast, encompassing security, advertising, and property rights protection, among others. For example, merchants or enterprises can employ this application to embed ``subtle'' advertisements within products without disturbing consumers. To support these functionalities, specific recognition algorithms are required according to different using purposes.
In Figure \ref{fig:app} (d), we provide a sketch of an implementation where we generated and printed an icon (a computer icon on a piece of paper, composed of numerous imperceptible QR codes. Aesthetically, it appears not much different from standard printed content. Touching the phone to the printed icon activates certain actions, such as turning on a PC. Ultimately, different QR codes, barcodes, and the like can be generated for various purposes. In the provided sketch demo, we employed a QR code for simplicity, which could be replaced with custom encoding, such as Anoto's dot pattern.

\subsubsection{Enhanced Ambient Camouflage}

Blending smartphones with their surroundings as if they are imperceptible and seamlessly integrated into the background may offer enhanced aesthetics and social discretion. For instance, during an intimate dinner, a phone on the table could camouflage itself \cite{10.1145/3025453.3025482} and blend into the background, thereby improving the user experience and promoting social acceptance. This level of realism appears increasingly feasible as edge-to-edge phones with minimal bezels and under-display cameras become more commonplace. Detailed images captured by microscope cameras can provide a more sophisticated foundation for imitation in this application, allowing for the incorporation of richer texture details to augment the verisimilitude of the camouflage images.
This application can be employed in social situations, meetings, and scenarios that necessitate discretion. In addition to the surface image captured by the MicroCam itself, some supplementary techniques such as edge alignment and image rendering are required to integrate local detail textures into the overall image. Moreover, to achieve superior real-time rendering effects, additional hardware, such as a high-performance cloud server, may be necessary. In our example in Figure \ref{fig:app} (e), for the purpose of sketch demonstration, the image captured by a conventional camera is displayed on the phone and manually adjusted (e.g., zooming in/out, alignment) to achieve the optimal camouflage angle. Subsequently, we employed manual photo-editing techniques to integrate the texture patterns of the microscopic images into photos of wooden materials, captured via conventional cameras. This procedure enhanced the texture details of the camouflage images, resulting in an augmented level of realism.

\subsection{Discussion 3: User Experience Analysis}

A plethora of intriguing and innovative sensing technologies have been consistently investigated for material and surface detection. However, limited research has concentrated on interaction factors beyond the sensor itself. In response, we strive to bridge these gaps by conducting a qualitative analysis of the positive and negative impacts of specific technical parameters on user experience. This approach will elucidate the differences between our contribution and prior work. Given the similarities in research context and technical direction, we have selected MagicFinger~\cite{yang2012magic}, SpectroPhone~\cite{SpectroPhone}, and SpeCam~\cite{Yeospecam} for comparative analysis.

As demonstrated in Table \ref{tab:comparison}, both MicroCam and MagicFinger perform material classification directly based on microscopic images. The key distinction lies in MicroCam's processing of RGB images, while MagicFinger utilizes grayscale images. The advantages of employing RGB images are evident; for instance, in section \ref{study1}, we assessed that classification results derived from RGB microscopic images surpass those obtained from grayscale images. This is attributable to the fact that color information is instrumental in distinguishing material textures with similar structures, which are lost in grayscale images. A compelling example involves the samples from objects cabinets (fiberboard material) and sinks (ceramic material), which are texturally similar and indistinguishable, yet their colors differ significantly. Moreover, we observed that our MicroCam possesses a higher resolution than MagicFinger, which proves beneficial for more precise context perception and broader application scenarios.

Moreover, both MicroCam and SpectroPhone utilize the rear camera instead of the front one. SpeCam requires the user to place the mobile phone face down, occupying the front display to showcase different lights, rendering the phone unusable during the sensing period. This approach is evidently less user-friendly and more challenging to use.
Additionally, the focal length, or sensing distance, of MicroCam is approximately 1mm. This necessitates more support to maintain the requisite distance between the mobile phone and the surface, compared to other methodologies. However, MicroCam enables a more user-friendly solution by necessitating a less cumbersome setup. For instance, while SpectroPhone and SpeCam both mandate a custom but heavy mobile phone case, we merely requires a commonplace and lightweight 1mm commercial mobile phone case or the simple use of a coin for elevation. This design choice notably enhances the user experience.
Ultimately, compared to other technologies, our MicroCam offers more detailed and fine-grained microscopic images, providing a broader range of context-aware information. For instance, multispectral imaging struggles to detect subtle differences within the same material, such as variations in surface texture. In contrast, our system can intuitively and easily discern these differences through microscopic images, as demonstrated by textiles with varying knitting densities in Figure \ref{fig:material}. Additionally, for the potential application example provided in section \ref{customizable_tag}, the multispectral-based method cannot identify specific surface textures such as ultra-small QR codes, while MicroCam's microscopic image can capture and display them intuitively.

Last but not least, in the final row of Table \ref{tab:comparison}, both MagicFinger and SpectroPhone utilize external, custom-built hardware to support their approaches. For instance, SpectroPhone requires external LEDs to provide a more personalized light source. In contrast, MicroCam and SpeCam depend solely on off-the-shelf components found in commercial cell phones.
While external hardware-based approaches have demonstrated feasibility, we focus more on the ``user experience in the present''. The development, commercialization, and public acceptance of a technology require time and resources. Our hardware solution has been applied to some commercial mobile phones, such as the Oppo Find X3 Pro and Realme GT2 Pro. With these devices, our method can be directly deployed on existing mobile phones without incurring additional hardware costs. For other mobile phones, users can also easily purchase various commercial microscope lens clips. Conversely, SpectroPhone's highly customized Bluetooth transmitter and receiver modules and LED lighting modules are currently not readily available on the market for users. Consequently, we contend that MicroCam offers distinct advantages in terms of practicality and convenience for user experience.

\begin{table}[!t]
  \centering
  \caption{Technical parameter comparison between 3 surface and material sensing methods (MicroCam, MagicFinger~\cite{yang2012magic}, SpectroPhone~\cite{SpectroPhone} and SpeCam~\cite{Yeospecam}). Note: numerical specifications outside resolution brackets represent cropped resolution, while figures within brackets indicate maximum resolution. For instance, 175*175 (248*248) means the maximum image resolution is 248*248 pixels and the cropped image resolution for algorithm processing is 175*175 pixels.} ~\label{tab:comparison}
  \resizebox{\textwidth}{!}{
  \begin{tabular}{l |c c c c}
    {\small\textit{Technical Parameters}}
    & {\small \textit{MicroCam}}
    & {\small \textit{MagicFinger}}
      & {\small \textit{SpectroPhone}}
       & {\small \textit{SpeCam}} \\
    \midrule

    Sensing Principle & RGB microscopic image & grayscale microscopic image & multispectral feature & multispectral feature\\
    Resolution (pixels) & 224*224 (1920*1080)  & 175*175 (248*248) & 640*480 & 3986*2976\\
    Sensing Distance(mm) & 1 & <5 & 3 & 3\\
    Front/Rear Camera & Rear &Tied to One Finger & Rear & Front \\
    Internal/External Sensor & all internal & all external & external LEDs+interal camera & all internal \\
  \end{tabular}
  }
\end{table}

\subsection{Discussion 4: Privacy Concern}

We realize that camera-based sensing approaches, such as MicroCam, may raise potential privacy concerns for users. On one hand, compared to other camera sensing methods that capture larger scenes, MicroCam focuses primarily on surface detail information rather than the overall scene. This makes it difficult to infer macro information, such as an individual's identity or building structure. From this perspective, the privacy risk associated with microscope camera-based sensing is lower than that of conventional camera-based sensing methods, offering certain advantages.
On the other hand, previous work has demonstrated that integrating cameras into various devices can mitigate some privacy and security issues \cite{padilla2015visual}. For instance, some device manufacturers provide physical kill switches for laptop cameras. Similarly, in terms of future enhancements, we anticipate the incorporation of a ``switching'' mechanism to regulate the camera's functionality within the MicroCam system. This could potentially manifest as a physical switch integrated directly into the mobile device or a virtual switch implemented through software algorithms. For the latter, it may necessitate the amalgamation of and support from authentication algorithms to ensure effective on/off control.
In addition, another possibility involves employing a low-resolution sensor. As demonstrated in the section \ref{study3}, the incorporation of continuous learning can enhance algorithm robustness and achieve better recognition results for challenging samples. Consequently, a lower-resolution sensor may increase the difficulty of reconstructing user information, thereby providing a degree of privacy protection.
In conclusion, while some current approaches address the inherent privacy concerns of camera sensing methods, this remains an ongoing challenge that requires further research and development to ensure user privacy is adequately protected.

\section{Limitations and Future Work}\label{future_work}
In our dataset, we included only 6 common objects and 9 typical materials encountered in our everyday life. 
However, the diversity of objects and materials in the wild can be far more extensive.
Therefore, in future work, we will consider more complicated situations. Additionally, we also plan to invite more users to further increase the dimension of the data, which is absolutely beneficial for future user-personalized context adaptation.

In the implementation of certain components within the system, we employ basic methods or algorithms. For instance, to detect the horizontal static state of a mobile phone, we simply compare the two data points obtained from the Inertial Measurement Unit (IMU), which are linear acceleration and angular acceleration, with a fixed threshold to determine the state. While this rudimentary approach proves to be highly effective, it is not without its limitations. The horizontal static state may trigger image sampling in some cases, which might not align with user expectations. To address this issue, we plan to consider the integration of additional sensors in future developments to enhance the system's performance in this regard. Additionally, we implemented our classifier using a high-end PC. Yet, it is possible to run the classifier on the phone itself, in real-time, since we are using a lightweight neural network architecture (i.e. MobileNet). Further, although we added a verification phase to check the ``object and material mapping relationship'' at the end of the system pipeline (as shown in Figure \ref{fig:MicroCam}), it still cannot exclude some special error cases. For example, as mentioned earlier, we noticed that from Table \ref{tab:mapping_relationship}, the object ``desk'' and ``cabinet'' contain two materials, ``particleboard'' and ``wood/wood-like grain''. Imagine for a moment, a wooden cabinet, where the object recognition result output by the model is incorrect (``table''), and the material recognition result is correct (``wood''), which will not be detected by our model. Therefore, in the future, we will try to incorporate data from multiple sensors (such as radar, etc.) into the system to adapt our prototype to more complex situations. Finally, the ER (Experience Replay) method utilized in the system optimization component represents a fundamental approach within the realm of continual learning. Our implementation of ER has indeed enhanced the system's robustness and generalization capabilities to a certain degree; however, there remains substantial potential for improvement. Continual learning is a dynamic and highly challenging field, offering numerous opportunities for further refinement and advancements in future research endeavors.


While macro cameras are becoming more common in smartphones today, not all devices feature a macro (microscope) camera. Therefore, we cannot claim that our sensing method will work on \emph{all} smartphones. In addition, we require a phone case that raises the gap between the camera and the surface by approximately 1mm to achieve optimal focus. However, with the addition of an external microscope lens (such as a clip-on) and a suitable phone case designed to fit over the standard phone camera, we suggest that MicroCam may perform well.


At present, we have only explored a limited number of applications. In future work, we aim to explore additional applications, such as continuous movement tracking of the mobile phone on a surface with unique textures (e.g., a wooden desk), akin to using a computer mouse. By integrating more information, such as additional sensor data or activity tracking within the software, we can further improve the intelligence and sophistication of MicroCam's context inference capabilities.

\section{Conclusion}

We have introduced a method for contact-based surface sensing using the built-in macro (microscope) camera of a smartphone. In our approach, IMU-based detection of a horizontal static state identifies users' actions when they naturally place the phone on any surface, anyway. Then, automatic surface recognition occurs via MobileNet-based microscopic image classification, further triggering corresponding actions such as placement-aware mode switching or body shortcuts.
We collected a substantial dataset, which supported us in training a robust neural network model capable of recognizing different surfaces with a high degree of accuracy. Moreover, we applied continual learning to optimize the robustness and generalization of the algorithm.
In conclusion, we anticipate that this sensing technique could, in due course, be available on any smartphone. The only requirement is a macro-lens feature, which is already emerging in the consumer market.

\bibliographystyle{ACM-Reference-Format}
\bibliography{sample}


\end{document}